\documentclass[]{article}

\usepackage[dvipsnames]{xcolor}
\usepackage{graphicx}
\usepackage{amssymb,amsmath,amsfonts}
\usepackage{todonotes}
\usepackage{enumerate}
\usepackage{siunitx}
\usepackage{hyperref}
\usepackage[capitalise]{cleveref}
\usepackage[ruled,vlined]{algorithm2e}
\usepackage{enumerate}

\usepackage{authblk}

\newenvironment{pseudocode}[1][htb]
{% Update algorithm name
    \begin{algorithm}[#1]%
    }{\end{algorithm}}

\newcommand*\ttvar[1]{\texttt{\expandafter\dottvar\detokenize{#1}\relax}}
\newcommand*\dottvar[1]{\ifx\relax#1\else
    \expandafter\ifx\string,#1\string,\allowbreak\else#1\fi
    \expandafter\dottvar\fi}

\newcommand{\pa}{\partial}
\newcommand{\Dt}{\Delta t}

\SetCommentSty{mycommfont}

\colorlet{Mycolor1}{OliveGreen}

\newcommand{\ypa}[1]{{#1}}
\newcommand{\ypb}[1]{{#1}}
\newcommand{\ypc}[1]{{\color{black}{#1}}}
\newcommand{\ypd}[1]{{\leavevmode\color{black}{#1}}}
\newcommand{\ype}[1]{{\leavevmode\color{black}{#1}}}
\newcommand{\bluew}[1]{{\leavevmode\color{black}{#1}}}

\newcommand{\ya}[1]{{\color{black}{#1}}}
\newcommand{\yb}[1]{{\color{black}{#1}}}

\date{}
\title{Coarse-grained Stochastic Model of Myosin-Driven Vesicles into Dendritic Spines}
\author[1]{Youngmin Park\footnote{Corresponding author \href{mailto:ypark@brandeis.edu}{ypark@brandeis.edu}}}
\author[3]{Prashant Singh}
\author[1,2]{Thomas G. Fai}
\affil[1]{Department of Mathematics, Brandeis University, Waltham, MA 02453, USA}
\affil[2]{Volen Center for Complex Systems, Waltham, MA 02453, USA}
\affil[3]{International Centre for Theoretical Sciences, TIFR, Bengaluru 560089, India}

\begin{document}

\maketitle

%\tableofcontents

\begin{abstract}
 We study the dynamics of membrane vesicle motor transport into dendritic spines, which are bulbous intracellular compartments in neurons that play a key role in transmitting signals between neurons. We consider the stochastic analog of the vesicle transport model in [Park and Fai, The Dynamics of Vesicles Driven Into Closed Constrictions by Molecular Motors. Bull. Math. Biol. 82, 141 (2020)]. The stochastic version, which may be considered as an agent-based model, relies mostly on the action of individual myosin motors to produce vesicle motion. To aid in our analysis, we coarse-grain this agent-based model using a master equation combined with a partial differential equation describing the probability of local motor positions. We confirm through convergence studies that the coarse-graining captures the essential features of bistability in velocity (observed in experiments) and waiting-time distributions to switch between steady-state velocities. Interestingly, these results allow us to reformulate the translocation problem in terms of conditional mean first passage times for a run-and-tumble particle moving on a finite domain with absorbing boundaries at the two ends. We conclude by presenting numerical and analytical calculations of vesicle translocation.
\end{abstract}

\section{Introduction}

Pyramidal neurons, which make up roughly 80\% of the mammalian neocortex \cite{xu2016cultured}, receive tens of thousands of excitatory inputs that terminate on dendritic spines \cite{nimchinsky2002structure}. They are thought to serve essential functions related to normal brain function \cite{bloss2011evidence}, with defective spine formation implicated in Autism spectrum disorder and Alzheimer's disease \cite{penzes2011dendritic}. \ya{Several important questions remain open regarding the function of dendritic spine morphology and its homeostasis, including mechanisms of receptor transport to the postsynapse \cite{adrian2014barriers}, determinants of spine growth and atrophy \cite{hering2001dentritic}, and the relationship of spine morphology to neurological disorders such as fragile-X syndrome \cite{irwin2000dendritic}.}

Normal synaptic function requires intracellular transport for maintenance \cite{da2015positioning}: molecular motor proteins squeeze membrane vesicles known as recycling endosomes through submicron-sized spine necks to deliver surface proteins to the postsynaptic density. \ya{Recent technological advancements have allowed researchers to measure sub-micron phenomena including the dynamics of recycling endosomes inside dendritic spines \emph{in vivo} \cite{da2015positioning}. In addition to steady unidirectional motion into the spine, experimental observations include other types of vesicle motion including stalling in the spine neck (sometimes referred to as corking), and direction reversal leading to bidirectional movement \cite{park2006plasticity,wang2008myosin}. These observations indicate the possibility of multiple behaviors within the same cellular environment. This motivates the use of the mathematical language of nonlinear dynamical systems, which is commonly used to describe systems with multiple steady-states. Bifurcations are known to occur within these nonlinear dynamical systems, leading to dramatically different observed behaviors as the parameters are varied.}

\ya{Mechanistic biophysical modeling provides a probe of the bifurcations in vesicle motion and emergence of bidirectionality.}
We have previously explored the effects of constriction geometries on bistable velocities in a vesicle transport model \cite{park2020dynamics}. Our analysis revealed that long, thin spines tend to encourage unidirectional motion, whereas wide, stubby spines tend to allow bidirectional motion. This study used a mean-field model of vesicle trafficking and neglected noise, a common feature of molecular motors.

\ya{In order to capture this switching behavior, we develop a stochastic model of the motor-driven transport of vesicles into dendritic spines. Incorporating noise in this manner makes it possible to make biologically-relevant predictions such as the probability and time of vesicle translocation for a given constriction geometry.}

Studies of motor-driven transport in other contexts often include bidirectional changes in their models by incorporating a waiting time distribution or a probability per unit time to switch vesicle velocity \cite{muller2008tug,bressloff2009directed,newby2009directed,kunwar2011mechanical}. This model of bidirectional motion is often referred to as the \textit{tug-of-war} effect \cite{julicher1995cooperative,newby2010random,guerin2011motion,guerin2011bidirectional,allard2019bidirectional}.
\ya{Bidirectionality arises in our model because of competing motor species that pull the vesicle cargo in opposite directions. The inclusion of competing motor species reflects the fact that multiple species of myosin (e.g.~Myosin V and Myosin VI \cite{da2015positioning}) are found in dendritic spines. These different species of myosin walk in different directions along actin filaments. In particular, whereas Myosin V walks toward the plus end of actin filaments, Myosin VI walks toward the minus end \cite{walter2012myosin}.}

\ya{To study the timescale of switching between different metastable velocities and its influence on the rate of vesicle translocation, we first construct an \emph{agent-based model} in which each motor is simulated explicitly with attachment and detachment kinetics. Although the agent-based model is relatively straightforward to implement and allows us to track and control the microscopic properties of molecular motors, it is well-known that agent-based models are not typically amenable to mathematical analysis \cite{an2017optimization} and that their computational cost becomes increasingly prohibitive as the motor number increases. We therefore coarse-grain the model at a mesoscopic level of description by combining a discrete master equation with a continuous population-level PDE through statistical sampling.  The resulting method preserves the underlying stochasticity of the agent-based model while improving the efficiency and tractability of our simulations.}

A key distinction between the present work and most previous studies, which neglect the effect of confinement on the viscous drag caused by the surrounding fluid, is that we use lubrication theory to model the increased drag arising from the confined geometry of the spine. Whereas most previous studies incorporating fluid friction \ypa{use Stokes' drag law \cite{guerin2011bidirectional,smith2018assessing,bovyn2021diffusion} or neglect drag altogether \cite{allard2019bidirectional}, this approximation is expected to become inaccurate for translocation through narrow spine necks that subject the vesicle to a high degree of confinement}.

\ya{Our model allows for a significant conceptual simplification of vesicle dynamics and provides quantitative predictions on how the geometry (i.e. length and initial vesicle position) of dendritic spines influences the probability of vesicle translocation and the corresponding timescale related to the delivery of membrane receptors to the postsynaptic density. This delivery timescale has been implicated in synaptic growth and homeostasis \cite{bowen2017golgi}.}

\subsection{Idealized Dendritic Spines and Viscous Drag}

\begin{figure}[ht!]
    \centering
    \includegraphics[width=.8\textwidth]{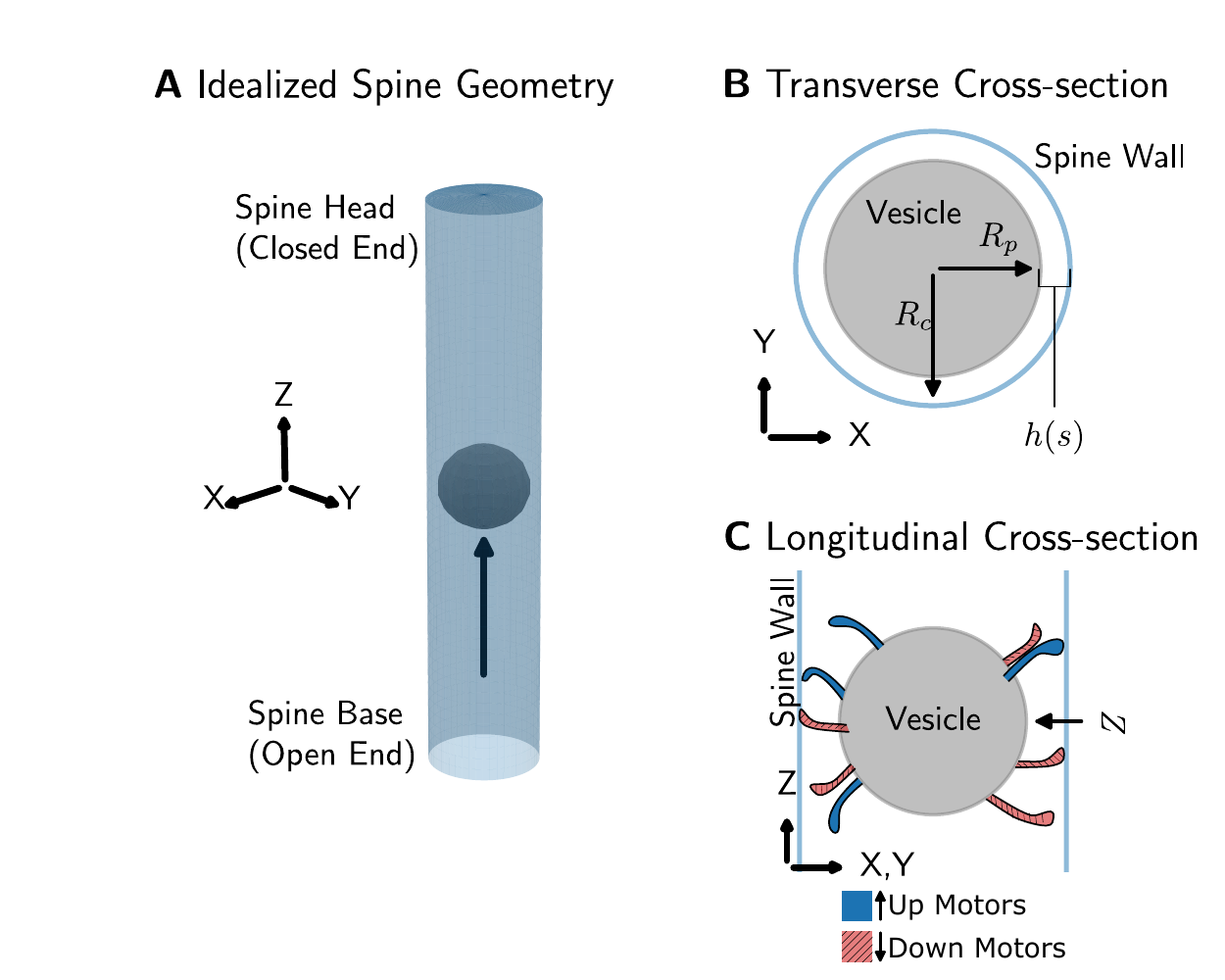}
    \caption{Idealized dendritic spine geometry. A: We consider idealized cylindrical spines (blue). The vesicle (gray sphere) initializes at the base (open end) and is shown traversing upwards towards the spine head (closed end) as indicated by the arrow. B: The spherical vesicle of radius $R_p$ traverses a spine neck of radius $R_c$, with $h(s)$ denoting the height of the vesicle surface above the spine wall in the center-of-mass frame $s \in [-R_p,R_p]$. C: Longitudinal cross-section of the spine, with the vesicle center of mass denoted by $Z$. \ya{Blue: up motors. Red: down motors.}}\label{fig:geometry}
\end{figure}

Dendritic spines protrude \SIrange{0.5}{1}{\um} \cite{risher2014} from the dendritic shaft and often appear mushroom-like in shape. Spines exhibit a thin \SIrange{50}{100}{\nm} \cite{harris1989dendritic} neck that opens into a head that is up to several hundred nanometers greater in diameter \cite{miermans2017biophysical}. To simplify the diverse morphology of dendritic spines \cite{yuste2010dendritic}, we consider a cylinder with an open end representing the base of the spine and a closed end representing the head of the spine (Figure \ref{fig:geometry}A). We assume the vesicle is a rigid sphere within a densely packed and highly viscous intracellular environment, resulting in overdamped vesicle motion.

First, we review the mean-field model of vesicle motion through dendritic spines that is the starting point for the present work. Assuming the ratio of vesicle diameter to cylinder diameter is close to unity, results from lubrication theory apply \cite{acheson1991elementary,fai2017active,park2020dynamics}, yielding the deterministic system of ordinary differential equations
\begin{subequations}
    \begin{align}
        \frac{dZ}{dt} &= V, \label{eq:lubrication_eps1}\\
        M \frac{dV}{dt} &= \bar F_\text{motors}(V) - \zeta(Z) V,\label{eq:lubrication_eps2}
    \end{align}
\end{subequations}
with initial conditions $Z(0) = Z_0$ and $V(0) = V_0$. $\bar F_\text{motors}(V)$ represents the mean-field molecular motor force output, $V$ is the vesicle velocity, $Z$ is the vesicle position, $M$ is the vesicle mass, and $\zeta(Z)$, defined as,
\begin{equation}\label{eq:zeta}
    \ypd{\zeta(Z) = \int_{-R_p}^{R_p} \left[R_ph^{-2}(Z+s) + R_c^2h^{-3}(Z+s) \right]ds,}
\end{equation}
is a drag coefficient that arises due to the constriction geometry. The function $h(Z+s) = \ypd{\tilde R_c(Z+s) - \sqrt{R_p^2-{s}^2}}$ \ypd{for $s \in [-R_p,R_p]$} is the distance from the vesicle to the spine wall given the vesicle center of mass $Z$ (Figure \ref{fig:geometry}B,C), $\tilde R_c(z)$ is the constriction radius as a function of position $z$, $R_c = \min_z \tilde R_c(z)$ is the minimum constriction radius, and $R_p$ is the vesicle radius. \ypc{Note that the height function $h$ is constant in our problem because $\tilde R_c(Z+s)$ is constant in $Z$ and therefore $\zeta(Z)$ is constant in $Z$. For a detailed derivation of Equations \eqref{eq:lubrication_eps1} and \eqref{eq:lubrication_eps2} in the general case where $\zeta$ depends on $Z$, we refer the reader to \cite{fai2017active,park2020dynamics}}.

The motor force appearing on the right-hand side of \eqref{eq:lubrication_eps2} is a nonlinear deterministic function of the velocity\ypb{; its explicit functional form is given in Section \ref{sec:lang}}. While this force-velocity curve is known to capture the empirical behavior of molecular motors at a population level and leads to multistability \ypb{\cite{park2020dynamics}}, it is unable to capture the stochasticity \ypd{in velocity observed in experiments \cite{park2006plasticity,wang2008myosin}}. In particular, on the order of a few dozen molecular motors are thought to be present in dendritic spines, so fluctuations in motor dynamics may play an important role in bidirectional vesicle motion. \bluew{To account for these effects, we develop a stochastic version of \eqref{eq:lubrication_eps2} based on the influence of individual myosin motors (i.e. agents) on vesicle motion. Unlike the mean field model, the stochastic version correctly captures velocity fluctuations and also predicts the distribution of waiting times to transition between quasi-steady states. This model also allows us to recast the problem of vesicle translocation in terms of the dynamics of a run-and-tumble particle.}

Before introducing the stochastic model, we first take note of important features of the mean-field model relevant to the present study. First, in contrast to our prior work in \cite{park2020dynamics}, here for simplicity we assume that the spine wall does not change in diameter as a function of the vesicle center of mass $Z$. Therefore, $\zeta$ is constant in $Z$. Second, the mean-field force-balance equation in the overdamped limit $M \rightarrow 0$ may be taken directly from Equation \ypa{\eqref{eq:lubrication_eps2}}:
\begin{equation}\label{eq:balance}
    0 = \bar F_\text{motors}(V) - \zeta V.
\end{equation}
That is, the total motor force $\bar F_\text{motors}(V)$ must balance the drag effects due to confinement $\zeta V$, and this force-balance is instantaneous relative to the vesicle velocity and motor dynamics. \ypd{Nonzero velocities satisfying this force-balance equation are the quasi-steady state velocities, denoted $\pm V^*$.} As explained in Section \ref{sec:vesicle_dynamics}, we use a similar form of this force-balance equation when simulating the force output of individual molecular motors in the agent-based model.  Finally, instead of taking the mean-field force $\bar  F_\text{motors}(V)$ as in our previous work, here we consider an explicit sum of discrete myosin motor forces. \bluew{The precise form of $\bar  F_\text{motors}(V)$ is introduced in Section \ref{sec:lang}.}

This paper is organized as follows. In Section \ref{sec:agents}, we describe the agent-based model of myosin motors and provide several numerical examples of the tug-of-war effect. In Section \ref{sec:coarse_grain}, we coarse-grain the agent-based model using a Langevin approximation, which fails to capture the agent-based model dynamics. We then coarse-grain using a master equation, which accurately captures the desired dynamics. Finally, in Section \ref{sec:translocation} we use the coarse-grained master equation to compute vesicle translocation times and probabilities. We further simplify the translocation problem by viewing the vesicle as a telegraph process.

An open-source repository is available on GitHub at \url{https://github.com/youngmp/NoisyMotors}.

\section{Agent-Based Myosin Motor Model}\label{sec:agents}

There is strong experimental evidence that the actin-myosin cytoskeleton dominates transport into dendritic spines \cite{da2015positioning}. In order to better understand the biophysical principles underlying this behavior, we develop a model for vesicle transport based on the action of individual myosin motors. The agent-based model has also been considered in the context of skeletal muscle \cite{huxley1957muscle,lacker1986mathematical,hoppensteadt2012modeling}. Below, we introduce this model of the motor-driven transport of vesicles into dendritic spines.

\ya{Motors have an intrinsic polarity in the sense that they exert forces in a particular direction along microtubules. For example, myosin V walks toward the \yb{plus} end of actin. However, the cargo is not necessarily constrained to move in the same direction as the motors. This may occur, for instance, if there is a tug-of-war of competing motors with opposite polarities. The situation in which the motors pull with precisely equal and opposite forces is unstable---eventually only one of the motors will win out, and from the perspective of the losing motor the cargo will move opposite to its own direction of force production. We refer to these two alternatives by introducing the terminology of \emph{preferred} and \emph{non-preferred} directions of motion. In the case of competing motors, it is critical to account for this effect (which is sometimes referred to as a superisometric force \cite{hoppensteadt2012modeling} in the case of muscle myosin). These asymmetrical forces exerted in the preferred and non-preferred directions are responsible for the multistability that arises in the model.}

The agent-based model consists of two identical myosin motor species that prefer to push the vesicle in opposite directions. The motors that prefer to push \ya{the vesicle down} (down motors) are denoted $d_i$ for $i=1,...,N$, and motors that prefer to push \ya{the vesicle up} (up motors) are denoted $u_i$ for $i=1,...,N$ (the number of up motors and down motors are both assumed equal to $N$).

\ya{We assume the cargo domain of each molecular motor is anchored to the vesicle, whereas the head of each motor freely attaches \yb{or} detaches from the spine cortex. The motors drive vesicle motion in a direction determined by the collective forces exerted on the vesicle (the motors intermittently attach and detach to the spine cortex with rates that we will describe shortly)}. This model satisfies classic tug-of-war dynamics \cite{muller2008tug} in the vesicle velocity $V$: the vesicle rapidly settles around a quasi-stable velocity $-V^*<0$ before switching to another quasi-stable velocity $V^*>0$ (recall that both quasi-steady state velocities $\pm V^*$ satisfy the condition $\bar F_\text{motors}(V) - \zeta V = 0$). This process continues, switching back and forth between the two quasi-stable velocities (Figure \ref{fig:agents}A). 

\subsection{Motors in the Preferred Direction} \yb{To simplify describing the motor dynamics without loss of generality, we assume that the tug-of-war dynamics place the vesicle in a downward motion, i.e., negative vesicle velocity $V<0$, so that it fluctuates around the quasi-stable velocity $-V^*$. We now quantify the contribution of each motor species to vesicle motion starting with a particular down motor.}

\begin{figure}[ht!]
    \centering
    \includegraphics[width=.8\textwidth]{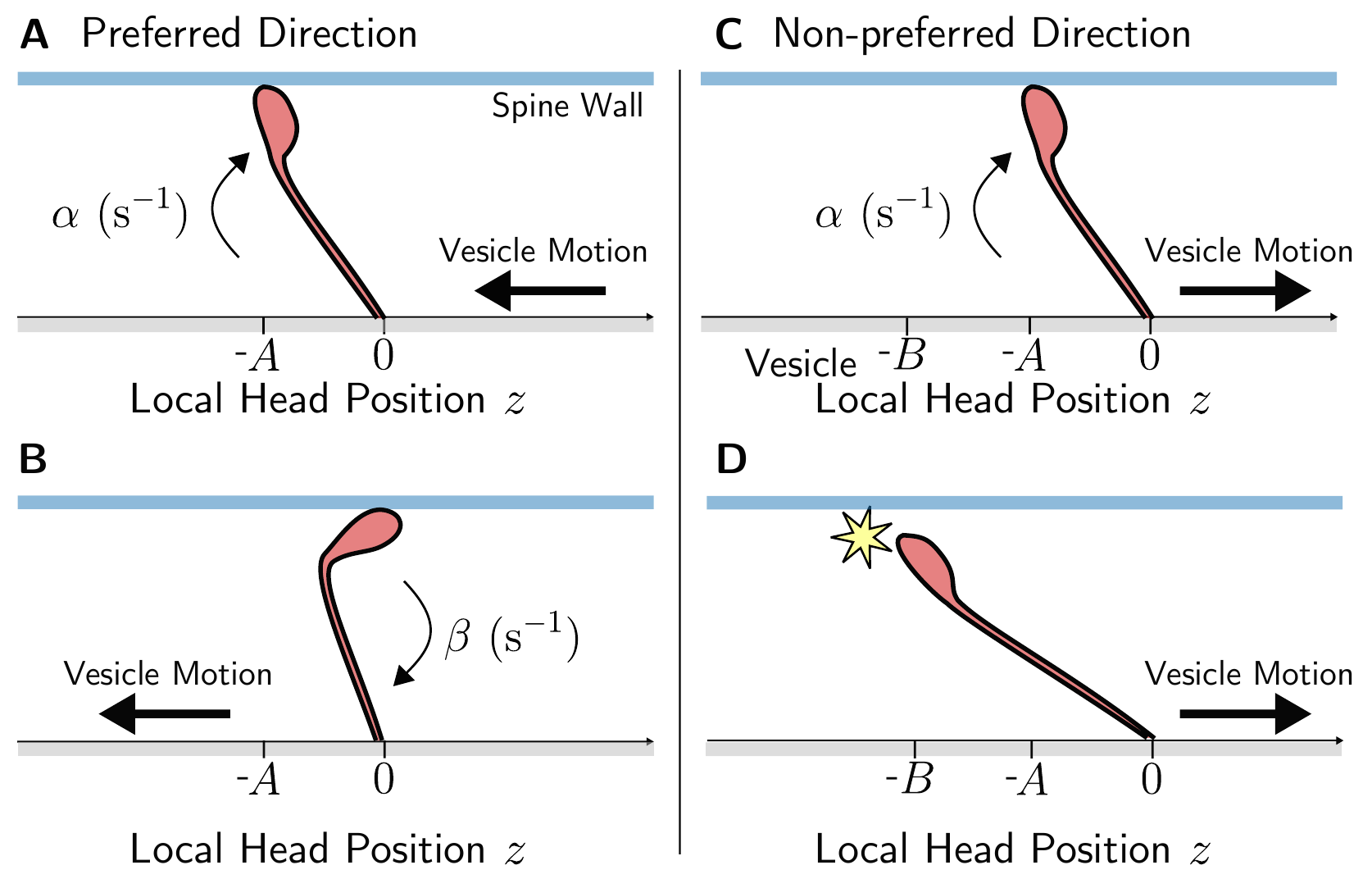}
    \caption{Individual myosin motor dynamics for down motors. Gray regions denote the vesicle surface, and blue regions denote the spine wall. The base of myosin motors is anchored to the \ya{vesicle}, and the head attaches and detaches on the \ya{spine wall}. Black arrows indicate vesicle motion. A, B: After a down motor attaches (at a rate $\alpha$) to a vesicle moving with negative velocity ($V<0$, preferred direction), the detachment rate is $\beta$. C, D: In contrast, after a down motor attaches to a vesicle moving with positive velocity ($V>0$, non-preferred direction), its detachment rate is position-dependent. In addition to the basal detachment rate $\beta$, extension beyond the yield position $z=\ya{-B}$ results in immediate detachment. The case of up motors is analogous, with the important difference of the opposite-sign preferred velocity.}\label{fig:myosin}
\end{figure}

Upon attachment, it experiences a velocity in its preferred direction ($V<0$). This motor attaches at a rate $\alpha$ and detaches at a rate $\beta$ (Figure \ref{fig:myosin}A,B). Its local head position, denoted $z_d^i$ (given relative to the location of the base and independent of the global vesicle position $Z$ in Figure \ref{fig:geometry}), satisfies the ODE:
\begin{equation}\label{eq:position_dynamics}
    \frac{dz_d^i}{dt} = \begin{cases}
        \yb{-V(\{z_u^j,z_d^k\})}, & \text{if attached}\\
        0, & \text{if unattached}
    \end{cases},
\end{equation}
\ya{with initial condition $z_d^i(0) = -A$}. \yb{The velocity depends on all motor positions $\{z_u^j,z_d^k\}$ --- we later describe this dependence explicitly in Equation \eqref{eq:agents_fastv}}. Whenever this down motor detaches, it instantaneously returns to its rest position $z_d^i=0$. These dynamics hold for any down motor, and we calculate the total force output of down motors by taking the sum:
\begin{equation} \label{down-force-eq-1}
    F_d\ypd{(\{z_d^i\})} = \sum_{i=1}^{N}f\left(z_d^i\right),
\end{equation}
where $N$ denotes the total number of down motors, $z_d^i$ is the position of attached motor $i$, and in general, $f(z)$ is the force exerted by a single motor at position $z$, and $f(0)=0$. For example, $f(z)=p_1[\exp(\gamma z)-1]$ and $f(z) = p_1\gamma z$ are acceptable choices \cite{hoppensteadt2012modeling}, but we proceed with the linear function for simplicity. Note that the force output of the motor is position-dependent, in contrast to models of kinesin and dynein motors (cf. \cite{allard2019bidirectional}), and that detached motors with position $z=0$ exert no force because $f(0)=0$. \textit{This local position-dependence will be an important feature to consider when coarse-graining the agent-based model}.

\subsection{Motors in the Non-Preferred Direction}
Next, consider a particular up motor which moves in its non-preferred direction because the velocity $V<0$ is opposite of its preferred direction. The attachment rate is $\alpha$, just as in the case of down motors. However, upon attachment at extension \ya{$z_u^i=A$, this motor experiences movement in its non-preferred direction and different dynamics come into play. In particular, there is a yield position $z_u^i=B$, with $B>A$}, at which the motor is assumed to detach. The position dynamics satisfy

\begin{equation}\label{eq:position_dynamics2}
    \frac{dz_u^i}{dt} = \begin{cases}
        \yb{-V(\{z_u^j,z_d^k\})}, & \text{if attached and } \ya{z_u^i \leq B}\\
        0, & \text{otherwise}
    \end{cases},
\end{equation}
with initial condition \ya{$z_u^i(0) = A$}. Note again: \yb{first, we later describe the velocity-position relationship in Equation \eqref{eq:agents_fastv}, and second}, $z_u^i$ is a local coordinate distinct from the vesicle center of mass $Z$ in Figure \ref{fig:geometry} and from the local down motor positions $z_d^i$. To summarize, \ya{as the vesicle moves the downward direction, it stretches the motor away from its attachment position until the motor detaches either due to the basal detachment rate $\beta$, or because it is extended beyond position $z_u^i=B$} and instantaneously detaches (Figure \ref{fig:myosin}C,D). \bluew{Such position-dependent detachment has also been considered in \cite{hoppensteadt2012modeling,fai2017active}}. These dynamics \eqref{eq:position_dynamics2} hold for any up motor moving in its non-preferred direction, and we calculate the total force output of up motors by taking the sum:
\begin{equation} \label{up-force-eq-1}
    F_u\ypd{(\{z_u^i\})} = \sum_{i=1}^{N} f(z_u^i),
\end{equation}
where $N$ denotes the total number of up motors (we assume equal total numbers of up and down motors), and $z_u^i$ is the local position of the $i$th up motor. The force is position-dependent, thus local motor positions also play an important role during non-preferred vesicle motion.

%Finally, if the vesicle velocity switches from $V<0$ to $V>0$, then the down (up) motors experience motion in their non-preferred (preferred) direction and the local motor position dynamics of down (up) motors  switch to Equation \eqref{eq:position_dynamics2} with $-A$ and $-B$ in place of $A$ and $B$ (Equation \eqref{eq:position_dynamics} with $A$ and $B$ in place of $-A$ and $-B$).

\bluew{In what follows, we use the forces exerted by down and up motors, which are given in \eqref{down-force-eq-1} and \eqref{up-force-eq-1}, respectively, to write the equations governing the dynamics of vesicle motion.}

\subsection{Vesicle Dynamics}\label{sec:vesicle_dynamics}
Having described the forces generated by down motors, up motors, and viscous drag and hence their contributions to the force-balance equation \eqref{eq:lubrication_eps2}, we may now write down the microscopic non-dimensional analog of the mean-field equations from Equations \eqref{eq:lubrication_eps1}, \eqref{eq:lubrication_eps2} from \cite{park2020dynamics}:
\begin{subequations}
    \begin{align}
        \frac{dZ}{dt} &= V, \label{eq:agents_velocity1}\\
        M \frac{dV}{dt} &= F_u\bluew{(\{z_u^i\})} \ya{+} F_d\bluew{(\{z_d^i\})}- \zeta V,\label{eq:agents_velocity2}
    \end{align}
\end{subequations}
with initial conditions $Z(0) = Z_0$ and $V(0) = V_0$. The variable $Z\in [0,L]$ is the center of mass of the vesicle and $L$ is the length of the dendritic spine. \bluew{We emphasize that equations \eqref{eq:agents_velocity1} and \eqref{eq:agents_velocity2} are stochastic differential equations due to the stochastic nature of $z_i^d$ and $z_i^u$ (see \eqref{eq:position_dynamics} and \eqref{eq:position_dynamics2}). On the other hand, the mean field equations \eqref{eq:lubrication_eps1}, \eqref{eq:lubrication_eps2} are deterministic differential equations. We illustrate in Section \ref{sec:direction_reversal} and Figure \ref{fig:agents} that this stochasticity gives rise to switching between quasi-steady states that is otherwise absent in the mean field description.} 

\bluew{Returning to Equations \eqref{eq:agents_velocity1} and \eqref{eq:agents_velocity2},} we assume that force-balance is instantaneous relative to the local motor positions, i.e., we consider \eqref{eq:agents_velocity2} in the overdamped limit $M\rightarrow 0$. In this limit, we may solve for the instantaneous cargo velocity,
\begin{equation}\label{eq:agents_fastv}
    V = \left [ F_u(\{z_u ^i\}) \ya{+} F_d(\{z_d ^i\}) \right]/\zeta.
\end{equation}
\ypd{Note that we may solve for the instantaneous velocity because the forces depend directly on motor positions which are effectively constant relative to the molecular motor dynamics. Such a solution is not possible with the mean-field force-balance equation $\bar F_\text{motors}(V) - \zeta V = 0$, because the mean-field motor forces depend directly on velocity (local motor positions are averaged out and thus do not appear).}

\subsection{Direction Reversal}\label{sec:direction_reversal}
Consider a vesicle moving with velocity $V<0$ near the quasi-stable velocity $-V^*$. As the result of stochastic fluctuations, it is possible that more up motors momentarily attach than down motors. Then the up forces dominate and the vesicle velocity switches direction from $V<0$ to $V>0$, rapidly settling around the other quasi-stable velocity $V^*$. The down motors now experience movement in their non-preferred direction and thus include the detachment dynamics in \eqref{eq:position_dynamics2} with corresponding sign changes. On the other hand, the up motors experience movement in their preferred direction and thus experience pure attachment and detachment dynamics as in \eqref{eq:position_dynamics} with corresponding sign changes. The vesicle position equation \eqref{eq:agents_velocity1} and force-balance equation \eqref{eq:agents_fastv} remain the same.

%Mersenne twister \texttt{=2}
\begin{figure}[ht!]
    \includegraphics[width=\textwidth]{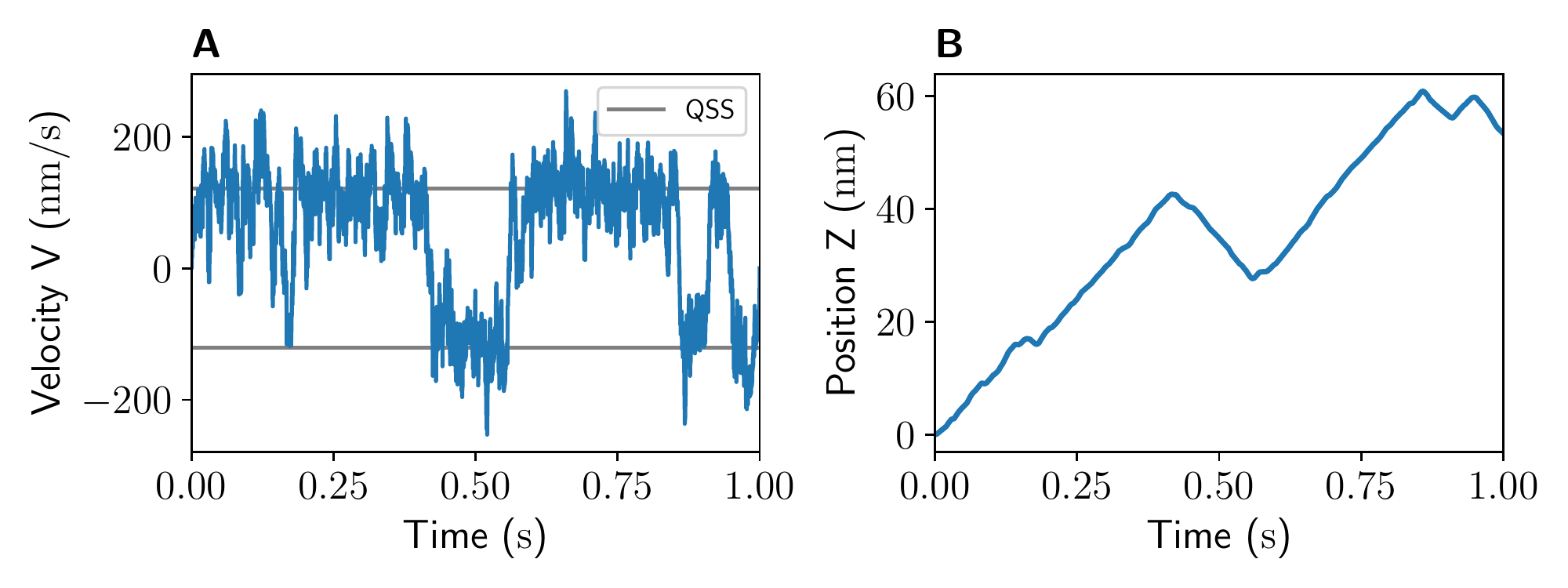}
    \caption{An example simulation of the agent-based model. A: Velocity (\si{\nm/\s}) over time (\si{\s}). Gray lines denote the quasi-steady state (QSS) velocities determined using a mean-field approximation \cite{park2020dynamics} of the agent-based model. B: Position (\si{\nm}) over time (\si{\s}). Model parameters: $N=100$, $\zeta=\SI{4e-5}{\kg/\s}$, $A=\SI{5}{\nm}$, $B=\SI{5.05}{\nm}$, $\alpha=\SI{14}{\s\tothe{-1}}$, $\beta=\SI{126}{\s\tothe{-1}}$, $p_1=\SI{4}{\pico\newton}$, $\gamma=\SI{0.322}{\nm\tothe{-1}}$. Simulation parameters: \texttt{dt=2e-6}, and \texttt{T=1} (we use \texttt{T} to denote the simulation end time throughout the text).}\label{fig:agents}
\end{figure}

We show an example of an agent-based model simulation in Figure \ref{fig:agents} using myosin motor parameters from \cite{hoppensteadt2012modeling,fai2017active}. The parameters are $A=\SI{5}{\nm}$, $B=\SI{5.05}{\nm}$, $\alpha=\SI{14}{\s\tothe{-1}}$, $\beta=\SI{126}{\s\tothe{-1}}$, $p_1=\SI{4}{\pico\newton}$, and $\gamma=\SI{0.322}{\nm\tothe{-1}}$. \ype{We choose the drag coefficient to be $\zeta=\SI{4e-5}{\kg/\s}$, which occurs when the constriction radius and vesicle radius have a ratio of approximately $\SI{1.5}{\um}/\SI{1}{\um}$ (this ratio is appropriate for stubby spines and results in nontrivial dynamics in the velocity, whereas ratios closer to unity may result in trivial, unidirectional motion \cite{park2020dynamics}). Given these parameters, solving for the quasi-stable velocity using the mean-field force-balance equation $0 = \bar F_\text{motors}(V) - \zeta V$ yields \bluew{$V = \pm V^*$} with $V^*=\SI{121}{\nm/s}$, which is close to the reported velocity range of myosin transport motors  \SIrange{200}{450}{\nm/\s} \cite{rief2000myosin}. In Section \ref{sec:translocation}  and Figure \ref{fig:mfpt_translocation} we consider translocation through spines of different lengths from \SIrange{500}{800}{\nm} \cite{majewska2000regulation}, where thin spines take approximately \SI{75}{\s} to translocate over \SI{1,200}{\nm} and stubby spines take approximately \SI{50}{\s} to translocate over \SI{750}{\nm} \cite{da2015positioning}.}

The velocity $V$ (\si{\nm/\s}) is shown in panel A and the position is shown in panel B with initial position $Z=\SI{0}{\nm}$, which represents the base of the dendritic spine. At the start of the simulation, the vesicle moves with an initial velocity of $V_0 = V^*=  \SI{121}{\nm/\s}$, which is the quasi-stable velocity predicted by the mean-field approximation \cite{park2020dynamics}. According to the agent-based model, the vesicle motion eventually undergoes a stochastic fluctuation and switches to a negative velocity and fluctuates about the other quasi-stable velocity $-V^* = \SI{-121}{\nm/\s}$. Appendix \ref{a:agents_euler} contains pseudocode for the agent-based model simulation.

Abstracting away from the individual motor dynamics, Figure \ref{fig:agents} shows that the essential features of the agent-based model include the existence of bistable velocities and the waiting-time distribution to switch between velocities. We prioritize these features because they determine the probability and conditional mean first passage time of a vesicle to successfully reach the other end of the spine (translocate). In the following sections, we seek to coarse-grain the agent-based model while preserving these essential features.

\section{Coarse-Graining the Agent-Based Model}\label{sec:coarse_grain}
The agent-based model is a helpful ground-truth model but inefficient to simulate. We first attempt to coarse-grain the agent-based model using the Langevin/Fokker-Planck approximation.

\subsection{Langevin Approximation}\label{sec:lang}

In this section, we approximate the agent-based dynamics using the Langevin equation,
\begin{subequations}
    \begin{align}
        \ypd{\frac{dZ}{dt}} &= V,\\
        dV &= \frac{(\bar F_\text{motors}(V)- \zeta V)}{M}dt + \frac{\sigma}{\sqrt{N}} dW \label{eq:langevin},
    \end{align}
\end{subequations}
with initial conditions $Z(0) = Z_0$ and $V(0) = V_0$. $\bar F_\text{motors}(V) = \bar F_u(V) \ya{+} \bar F_d(V)$, $V$ is the vesicle velocity, $\zeta$ is a viscous drag term,\bluew{ and $\bar F_d(V)$ and $\bar F_u(V)$ are the mean-field forms of forces $F_d\ypd{(\{z_d ^i\})}$ and $F_u\ypd{(\{z_u ^i\})}$, respectively [see \eqref{down-force-eq-1} and \eqref{up-force-eq-1}].} The force-velocity curve for down motors, $\bar F_{d}$, is
\begin{equation*}
    \bar F_{d}(V) = 
    \left\{
    \begin{array}{ll}
        \ya{-}\frac{p_1 \gamma N \alpha(V\ya{-}A\beta)}{\beta(\alpha+\beta)}  & \mbox{if } V \leq 0, \\
        \ya{-}\frac{p_1 \gamma N \alpha\left( V \ya{-} A\beta - e^{\beta(B-A)/V}(V \ya{-} B\beta)\right)}{\beta\left(\alpha-e^{\beta(B-A)/V}\alpha + \beta\right)} & \mbox{if } V > 0,
    \end{array}
    \right.
\end{equation*}
%\begin{equation*}
%	\bar F_{d}(V) = 
%	\left\{
%	\begin{array}{ll}
%		\ya{-}\frac{p_1 \gamma N \alpha(V+A\beta)}{\beta(\alpha+\beta)}  & \mbox{if } V \leq 0, \\
%		\ya{-}\frac{p_1 \gamma N \alpha\left( V + A\beta - e^{-\beta(B-A)/V}(V + B\beta)\right)}{\beta\left(\alpha-e^{-\beta(B-A)/V}\alpha + \beta\right)} & \mbox{if } V > 0,
%	\end{array}
%	\right.
%\end{equation*}
where $p_1$ (\si{pN}), $\gamma$ (\si{\nm^{-1}}) are parameters, $A$ is the local motor attachment position, $B$ is the local motor detachment position, $N$ is the number of down motors, and $\alpha$, $\beta$ are the attachment and detachment rates, respectively. This force-velocity curve can be derived using the same arguments in \cite{fai2017active} Section 3B, but with the linear force-position function $f(z) = -p_1 \gamma z$ in place of the force-position function $f(z) = -p_1[\exp(\gamma z)-1]$. Due to symmetry, the force-velocity curve for up motors, $\bar F_{u}$, is simply $\bar F_{u}(V)= \bar F_{d}(-V)$. As in the agent-based model, we assume the same number of $N$ up motors.

\ypa{Returning to Equation \eqref{eq:langevin}}, the term $dW$ is a white-noise process whose increments are normally distributed with mean 0 and variance $\Delta t$. We scale the white noise by the factor $ \sigma/\sqrt{N}$, where $N$ is the number of molecular motors of each species. Writing the noise magnitude in this form ensures that the standard deviation $\sigma$ converges to a constant value in the limit of large $N$, as observed in the agent-based model (Table \ref{tab:sigma}).

\begin{figure}[ht!]
    \includegraphics[width=\textwidth]{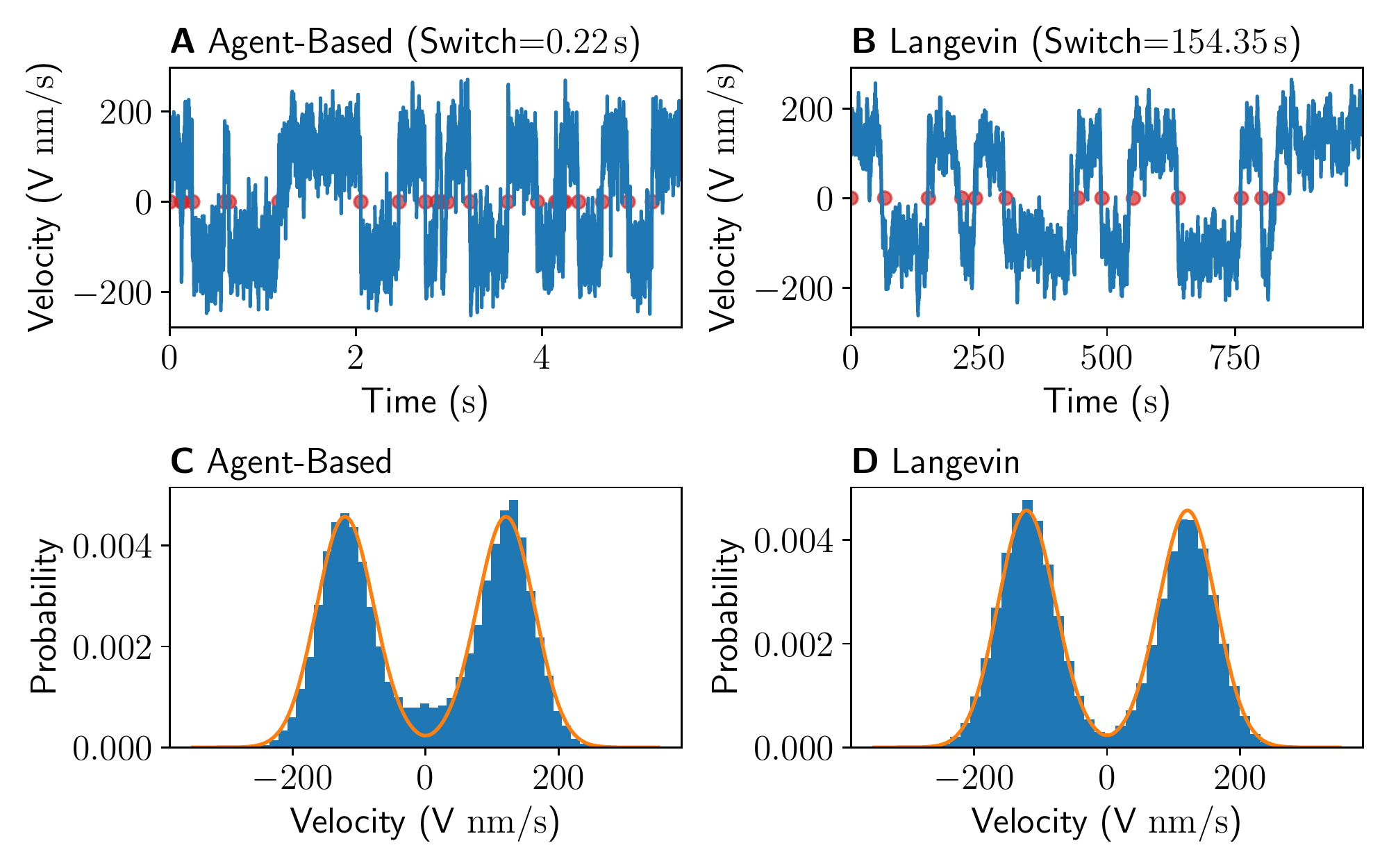}
    \caption{Steady-state distributions of the agent-based model and the Langevin equation. A: Velocity over time of the agent-based model. B: Velocity over time of the Langevin equation. Note the significant time difference in panes A and B. The agent-based model switches velocity at a significantly faster rate than the Langevin equation. A,B: Red dots indicate switching times. C: Steady-state distribution of velocities of the agent-based model. Orange curve represents the analytic steady-state distribution from the Langevin equation. D: Steady-state distribution of velocities in the Langevin equation.  C,D: Orange curves represent the analytic steady-state distribution from the Langevin equation. Parameters for the agent-based model are identical to Figure \ref{fig:agents}. Parameters for the Langevin equation are identical to Figure \ref{fig:agents} but with \ypb{$\sigma=\SI{423}{\nm^2/\s^2}$}.}\label{fig:langevin}
\end{figure}

\ypd{We may write the Fokker-Planck equation for the probability density function $p(V, t)$ corresponding to \eqref{eq:langevin} \cite{gardiner1985handbook}:
    \begin{equation*}
        \frac{\partial}{\partial t}p(V,t)=-\frac{\partial}{\partial V}\left(\frac{F_\text{motors}(V)-\zeta V}{M}p(V,t) \right)+\frac{\partial^2}{\partial V^2}\left(\frac{\sigma'^2}{2} p(V,t)\right).
    \end{equation*}
    This yields the following equation at steady-state assuming zero flux through the boundaries:
    \begin{equation*}
        \frac{\bar F_\text{motors}(V)- \zeta V}{M}p_s(V) - \frac{\left(\sigma'\right)^2}{2} \frac{d}{dV}[p_s(V)] = 0,
    \end{equation*}
    where $p_s$ is the steady-state distribution in vesicle velocity, and $\sigma ' = \sigma/\sqrt{N}$.
    This equation has the solution:}
\begin{equation}\label{eq:pss}
    \ypa{    p_s(V) = \mathcal{N}\exp\left[\frac{2}{(\sigma')^2}\int_{0}^V \frac{\bar F_\text{motors}(V')- \zeta V'}{M} dV'\right] },
\end{equation}
\bluew{where $\mathcal{N}$ is fixed by the normalization condition $\int_{-\infty}^{\infty}p_s(V)dV = 1$.}

To test whether a Langevin approximation correctly captures switching times, we simulated the agent-based model and plotted the bimodal steady-state distribution of velocities using $N=100$ motors (Figure \ref{fig:langevin}C). We then fitted the quantity $\sigma' := \sigma/\sqrt{N}$ to this steady-state distribution as described next.

\ya{To fit the noise magnitude $\sigma'$ we vary the number of motors and comparing the steady-state distribution of velocities to the probability density function given by \yb{\eqref{eq:pss}}. We verify that the noise scales in terms of the number of motors $N$ as $\sigma' = \sigma/\sqrt{N}$, as suggested by the central limit theorem, for some constant $\sigma$. This yields a general form that we use to define the noise magnitude independent of the number of motors.}

The resulting fit is shown as the orange curves in Figure \ref{fig:langevin}C,D. We show a distribution of velocities from the Langevin simulation in Figure \ref{fig:langevin}D. As expected, the simulation (blue bars) conforms to the analytically-computed steady-state distribution (orange curves).

\begin{table}[ht!]
    \begin{center}
        \caption{The fitted noise magnitude $\sigma'$ of the agent-based model \ypd{is given in units of \si{\nm/\s^{3/2}}} as a function of motor number $N$. Parameters are identical to Figure \ref{fig:langevin}, but to maintain the same total force as a function of motor number, we scale the parameter $p_1$ by a factor $q$ such that $q= 100/N$. Simulation parameters: \texttt{T=10}, \texttt{dt=5e-6}.}
        \label{tab:sigma}
        \def\arraystretch{2}%  1 is the default, change whatever you need
        \begin{tabular}{c|c|c|c|c|c} % <-- Alignments: 1st column left, 2nd middle and ♦3rd right, with vertical lines in between
            $N$ & 50 & 100 & 150 & 200 & 250\\
            \hline
            $\sigma'$ & 64.4 & 42.3 & 34.6 & 30.7 & 26.0 \\
            \hline
            $\sigma' \times \sqrt{N}$ & 443 & 423 & 412 & 420 & 411\\
        \end{tabular}
    \end{center}
\end{table}

The noise magnitude in the Langevin equation, $\sigma/\sqrt{N}$, scales as the reciprocal of the square root of the number of motors of each species, $1/\sqrt{N}$. We confirm that the same relationship holds in the agent-based model. Recalling that $\sigma'$ is the fitted noise magnitude in the agent-based model, multiplying $\sigma'$ by $\sqrt{N}$ should yield a constant value as a function of $N$. This property is shown in Table \ref{tab:sigma}: as we vary the number of motors in the agent-based model from $N = 50$ to $N = 250$, $\sigma'$ decreases (second row). Multiplying the second row by $\sqrt{N}$ yields an approximately constant standard deviation in the range \ypd{\SIrange{411}{443}{\nm/\s^{3/2}}} (third row). The Langevin equation appears to serve as a good approximation to the agent-based model because the steady-states agree and the noise magnitude of both models depend on $1/\sqrt{N}$.

\textit{However, despite this good agreement, an important property does not match}: the mean time to switch velocities differs significantly between the two models. On average, the agent-based model takes \SI{0.51}{\s} to switch velocity (Figure \ref{fig:langevin}A), while the Langevin equation takes \SI{117.44}{\s} to switch velocity. We conclude that the Langevin approximation of the agent-based model is not sufficient to capture agent-based dynamics. Indeed, in general, the Fokker-Planck equation (and therefore the Langevin equation) is a poor approximation for birth-death processes \cite{doering2005extinction}.

\subsection{Master Equation}

In this section, we turn to the next level of coarse-graining using the master equation. In contrast to the agent-based model, in which we check whether or not individual motors attach or detach at each time step, the master equation formulation only requires equations for the \textit{total} number of attached motors for each species. This formulation only requires knowledge of the growth and decay rates in the total number of attached down motors $D$ and the total number of attached up motors $U$.% Only the growth and decay rates are needed to simulate the master equation.

We first discuss these rates as a function of vesicle velocity. When the velocity is negative, $V<0$, the down motor population $D$ grows at a rate $\alpha(N-D)$ and decays at a rate $D\beta$. On the other hand, the up motor population $U$ grows at a rate $\alpha(N-U)$ and decays at a rate $\gamma_U(V) := U\beta/[1-\exp(-\beta(B-A)/|V|)]$. When the velocity is positive, $V>0$, the down motor population grows at a rate $\alpha(N-D)$ and decays at a rate $\gamma_D(V) := D\beta/[1-\exp(-\beta(B-A)/|V|)]$, while the up motor population grows at a rate $\alpha(N-U)$ and decays at a rate $U\beta$ (See Table \ref{tab:rates} for a full summary of attachment and detachment rates, and Appendix \ref{a:detachment_derivation} for the derivation of $\gamma_i(V)$).

When a motor population encounters its non-preferred velocity, the decay rate of the population depends on the vesicle velocity, which is coupled to the motor states by the force-balance equation,

\begin{equation}\label{eq:master_force}
    F_u(\{z_u^i\}) \ya{+} F_d(\{z_d^i\})- \zeta V = 0,
\end{equation}

\ypd{
    where the motor forces are given by the sum over motor positions:
    \begin{equation*}
        F_d\ypd{(\{z_d^i\})} = \sum_{i=1}^{N}f\left(z_d^i\right), \quad F_u\ypd{(\{z_u^i\})} = \sum_{i=1}^{N}f\left(z_u^i\right),
    \end{equation*}
    Note the implicit, fundamental difficulty with this equation, namely that it requires knowledge of motor positions. We do not wish to integrate individual motor positions using Equations \eqref{eq:position_dynamics}, \eqref{eq:position_dynamics2} or else we formulate an equally complicated version of the agent-based model}. In order to properly coarse-grain the agent-based model, we must derive an accurate approximation of motor positions over time without explicitly integrating every motor position.

\begin{table}[h!]
    \begin{center}
        \caption{Attachment and detachment rates in the master equation, with $\gamma_i(V)=i\beta/[1-\exp(-\beta(B-A)/|V|)]$. }
        \label{tab:rates}
        \def\arraystretch{2}%  1 is the default, change whatever you need
        \begin{tabular}{l|l|l} % <-- Alignments: 1st column left, 2nd middle and 3rd right, with vertical lines in between
            & Down $D$ & Up $U$\\
            \hline
            Attach Rate & $(N-D)\alpha$ & $(N-U)\alpha$ \\
            \hline
            Detach Rate $(V \leq 0)$ & $D\beta$ & $\gamma_U(V)$\\
            \hline
            Detach Rate $(V > 0)$ & $\gamma_D(V)$ & $U\beta$\\
        \end{tabular}
    \end{center}
\end{table}

\subsubsection{Motor Population Density}

\ypa{Let $\phi_D(z,t)$ denote the down motor position probability density, where $z$ is the local motor position and $t$ is time. The evolution of $\phi_D(z,t)$ is given by expressing the conservation of motor number through the following partial differential equation:
\begin{equation}\label{eq:advection}
    \frac{\pa}{\pa t}\phi_D\left(z,t\right) \ya{-} V\frac{\pa}{\pa z}\phi_D\left(z,t\right) = \beta \phi_D(z,t) + \alpha(1-\theta)\delta(z\ya{+}A),
\end{equation}
%\begin{equation}\label{eq:advection}
%	\frac{\pa}{\pa t}\phi_D\left(z,t\right) + V\frac{\pa}{\pa z}\phi_D\left(z,t\right) = -\beta \phi_D(z,t) + \alpha(1-\theta)\delta(z-A),
%\end{equation}
where $\theta=D/N$ is the proportion of attached down motors and $z$ is the local down motor coordinate. \ypc{We use the steady-state functions given the steady-state velocity $V^*$ as initial conditions.} Equation \eqref{eq:advection} is a time-dependent version of the steady-state motor distribution equation explored in \cite{fai2017active}.} The $z$-domain is given by \ya{$[-B,\infty)$ and the left boundary at $z=-B$ is absorbing}. The partial differential equation (PDE) for up motors uses the same formulation as above with the corresponding sign changes. Note that in this case the domain is given by \ya{$z \in (-\infty,B]$ and and the right boundary at $z=B$ is absorbing}.% \ypb{Note that in this case the domain is given by $z \in [-B,\infty)$ and and the left boundary at z=-B is absorbing.} 

\begin{figure}[ht!]
    \includegraphics[width=\textwidth]{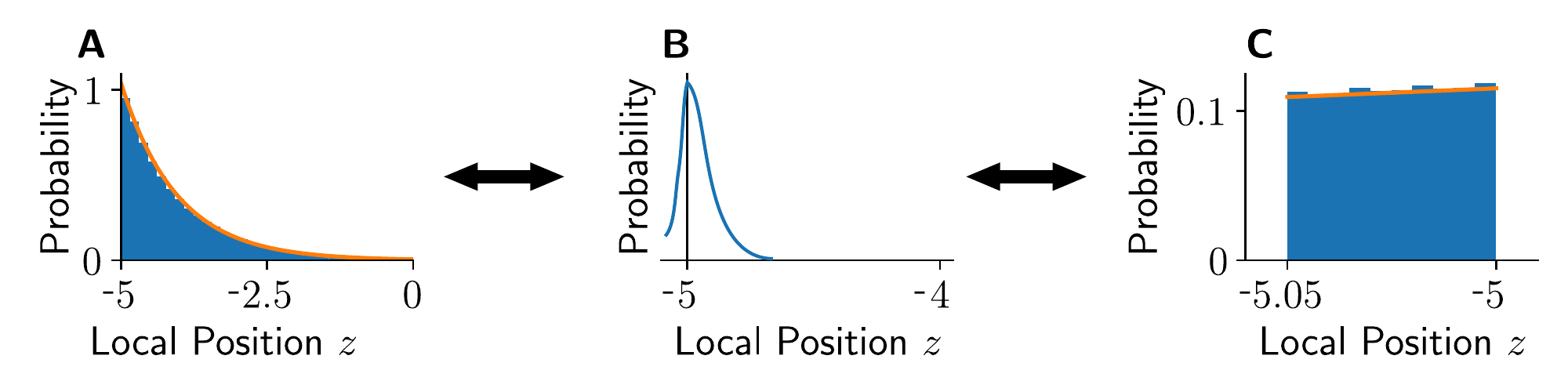}
    \caption{Probability density functions of down motor positions. Motors attach at \ya{$A=\SI{-5}{nm}$} and are dragged along by the vesicle. A: When the vesicle velocity $V$ fluctuates about the quasi-steady state $-V^*$, newly attached motors at \ya{$-A$} experience motion in their preferred velocity.  Blue bars indicate the numerically computed steady-state distribution and orange the analytically computed steady-state. B: Schematic of the probability distribution during a velocity switch from $-V^*$ to $V^*$. As the vesicle begins moving in the opposite direction, motors continue to attach at \ya{$z=-A$}, but \ya{attached motors with positions to the right of $z=-A$ begin to extend to the left of their attachment position, so the probability density now includes positions $z < -A$}. C: When the vesicle velocity $V$ fluctuates about the quasi-steady state $V^*$, down motors continue to experience vesicle motion in their non-preferred velocity. Any motors that extend beyond \ya{$B=\SI{-5.05}{nm}$} detach immediately. Orange curve and blue bars indicate the same quantities as in panel A. Behavior is identical but with opposite preferred-velocity in up motors \ya{and a different local position range of $(-\infty,B]$}). Model parameters identical to Figure \ref{fig:agents} but with a forced stationary velocity at $V=121$. Simulation parameters: \texttt{T=2}, \texttt{dt=1e-6}, \texttt{A0=0}.}\label{fig:distribution}
\end{figure}

We show representative solutions to this equation in Figure \ref{fig:distribution}. When $V=-V^*$, down motors attach at $z=\ya{-A}$ and the vesicle drags the motor heads towards less negative values of $z$ until they detach at a rate $\beta$. The distribution settles to a decaying exponential due to the memoryless detachment (Figure \ref{fig:distribution}A). When the velocity switches to the other quasi-steady state, $V^*$, down motors continue to attach at position $z=-A$ but are extended to increasingly negative values of $z$. While they continue to detach at a rate $\beta$, the yield position at \ya{$z=-B$} is captured by an absorbing boundary condition (Figure \ref{fig:distribution}C).

\subsubsection{Motor Forces and Closing the System}

\ypa{We preserve small number fluctuations in our coarse-grained approach by tracking individual motors and their attachment state, while using \eqref{eq:advection} to provide a statistical model of the extension dynamics of individual motors. That is, instead of tracking the extension lengths of individual motors, we sample lengths from the motor position probability density function.  Although \eqref{eq:advection} is deterministic, the resulting model includes stochasticity because of the possibility of fluctuations in the discrete numbers of attached motors. By solving this one-dimensional PDE to obtain a statistical model of extension lengths, we no longer need to track each motor position explicitly, thereby lowering the computational cost compared to the agent-based model.}

At each time step, we use the solution of Equation \eqref{eq:advection} as a probability distribution and draw local motor positions for each species. These positions are then put into the linear force-extension function and summed to produce the total force. Recall the force-balance equation:
\begin{equation*}
    F_u(\{z_u^i\}) \ya{+} F_d(\{z_d^i\}) - \zeta V=0.
\end{equation*}
We assume that force balance occurs on a much faster timescale than the motor population dynamics $D$ and $U$ and the underlying motor positions. We can then solve for the velocity $V$ based on the total number of motors attached and their positions \ypa{sampled from the solution to the PDE \eqref{eq:advection}}:
\begin{equation*}
    V = [F_u(\{z_u^i\}) \ya{+} F_d(\{z_d^i\})]/\zeta.
\end{equation*}

\subsubsection{Numerical Details of the Master Equation}

To solve the PDE of motor positions \eqref{eq:advection}, we use an upwinding scheme (Appendix \ref{a:upwinding}) on a non-uniformly spaced spatial mesh. One mesh partitions the interval $[-B,-A]$, and the other partitions the interval $[-A,A_0]$, where $A_0$ is chosen appropriately for the problem. While extensions may in principle allow motor positions to extend above $z=0$, this is negligible for our choice of parameters because the basal detachment rate $\beta$ is relatively large compared to the attachment rate $\alpha$. Therefore, we choose the upper boundary to be $A_0=0$.

Note that the motor distribution in the non-preferred direction (Figure \ref{fig:distribution}C) is on a significantly smaller domain compared to the preferred direction (Figure \ref{fig:distribution}A). We define the mesh such that the number of grid points $N_1$ above $z=-A$, equals the number of grid points below $z=-A$, $N_2$. Thus the spatial mesh size differs significantly above and below, i.e., above, $\Delta z_1 =(A_0-A)/N_1$, and below, $\Delta z_2 = (A-B)/N2$. $\Delta z_2$ is significantly smaller than $\Delta z_1$.

\begin{figure}[ht!]
    \includegraphics[width=\textwidth]{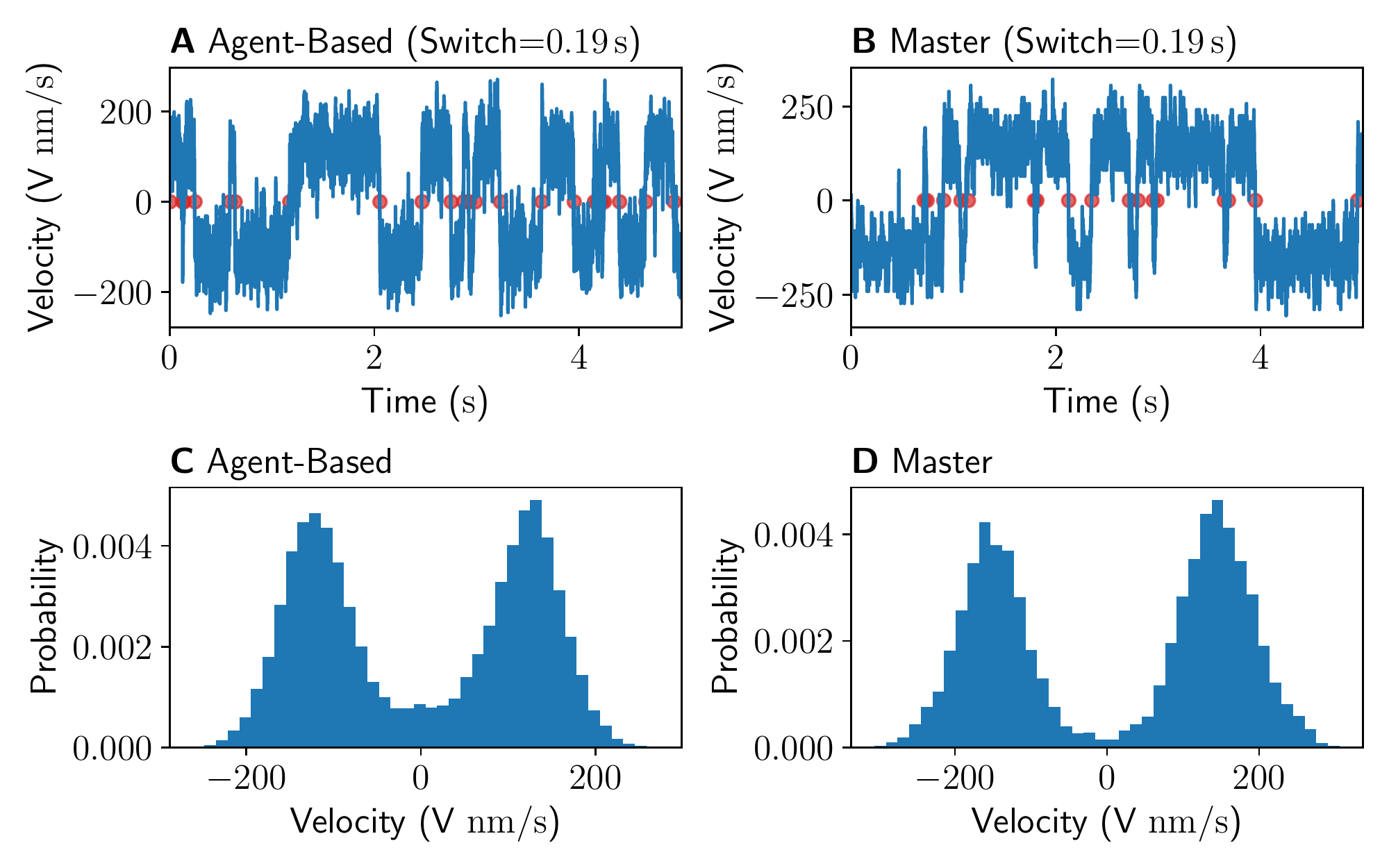}
    \caption{Velocity comparison between the agent-based model and master equation. In contrast to the Langevin equation, the master equation accurately captures the steady-state distribution and the mean time to switch velocity. Model parameters as in Figure \ref{fig:agents}. Agent-based model simulation parameters as in Figure \ref{fig:agents}. Master equation simulation parameters: \texttt{T=10}, \texttt{dt=3e-6}, \texttt{N1=N2=41}, \texttt{CFL=0.3}, \texttt{A0=0}.}\label{fig:master_v_agents}
\end{figure}

We define the Courant--Friedrichs--Lewy (CFL) condition based on the quasi-steady state velocity and the finer grid spacing below $z=-A$:
\begin{equation}\label{eq:cfl}
    CFL = \frac{V^* \Delta t}{\Delta z_2} \equiv V^* \Delta t \frac{N_2}{B-A}.
\end{equation}
We discuss the CFL using either grid points ($N_2$) or grid size ($\Delta z_2$). We typically choose $CFL \leq 0.3$ to ensure numerical stability --- the velocity fluctuates about the quasi-steady state velocity $V^*$, but virtually never by more than a factor of 3.

We show an example simulation comparing the master equation to the agent-based model in Figure \ref{fig:master_v_agents}. Panels A and C show the same velocity trace and probability density of the agent-based model as in Figure \ref{fig:agents}A,C. Panel B shows the velocity trace of the master equation, and panel D shows the corresponding distribution of vesicle velocities. We refer the reader to Appendix \ref{a:master_euler} for additional numerical details relating to the master equation, including pseudocode for simulations.

\begin{figure}[ht!]
    \centering
    \includegraphics[width=0.625\textwidth]{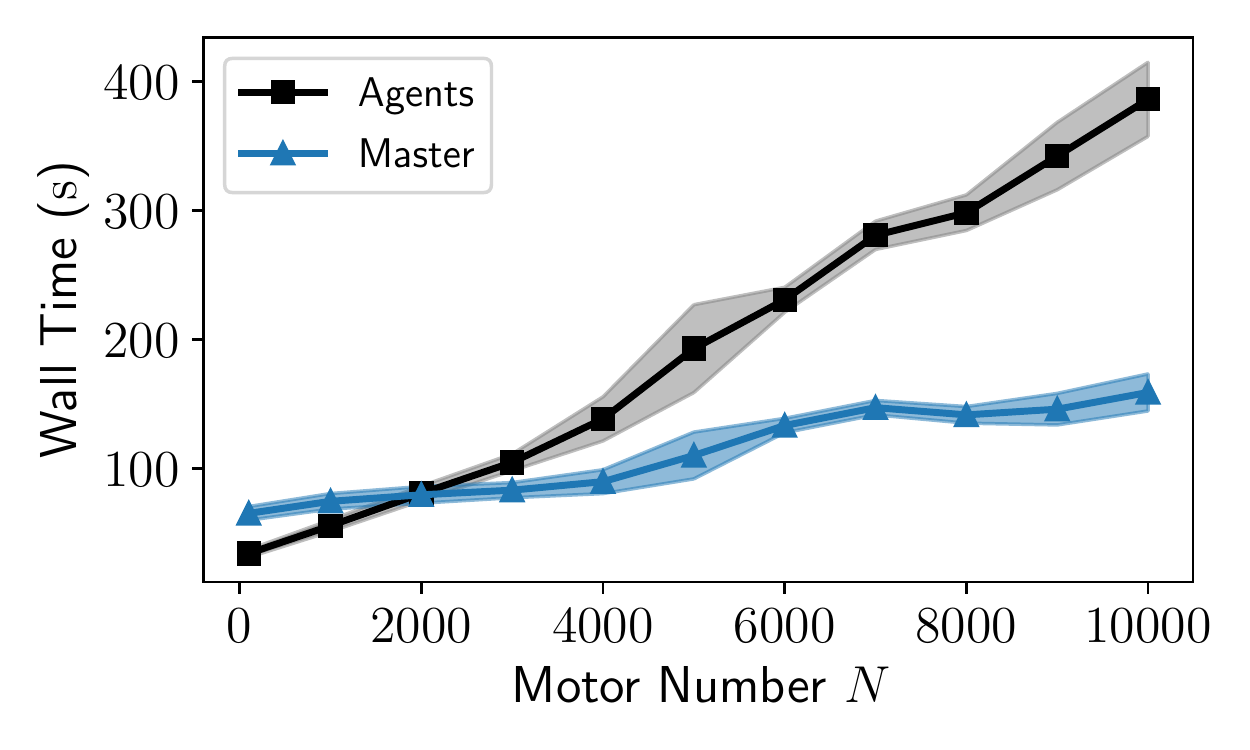}
    \caption{\ypa{Wall times of the agent-based model (black squares) and master equation (blue triangles). The master equation outperforms the agent-based model for greater numbers of motors. Model parameters as in Figure \ref{fig:agents} for both models. Agent-based model simulation parameters: \texttt{T=1}, \texttt{dt=3e-6}. Master equation simulation parameters: \texttt{T=1}, \texttt{dt=3e-6}, \texttt{CFL=0.3}, \texttt{N1=N2=41}.}}\label{fig:mva_times}
\end{figure}

\ypa{The master equation wall time performs within the same order of magnitude as the agent-based model with $N=100$ motors, but easily outperforms the agent-based model with $N=10,000$ motors (Figure \ref{fig:mva_times}). The significant performance difference arises from the differences in how the two models track motor positions and attachments. The agent-based model tracks the position of each motor as well as attachment and detachment states. In contrast, the master equation tracks only the total number of attached motors and draws the most likely motor positions from a probability distribution.}

\subsection{Mean Time to Switch Velocity}

\begin{figure}
    \centering
    \includegraphics[width=0.625\textwidth]{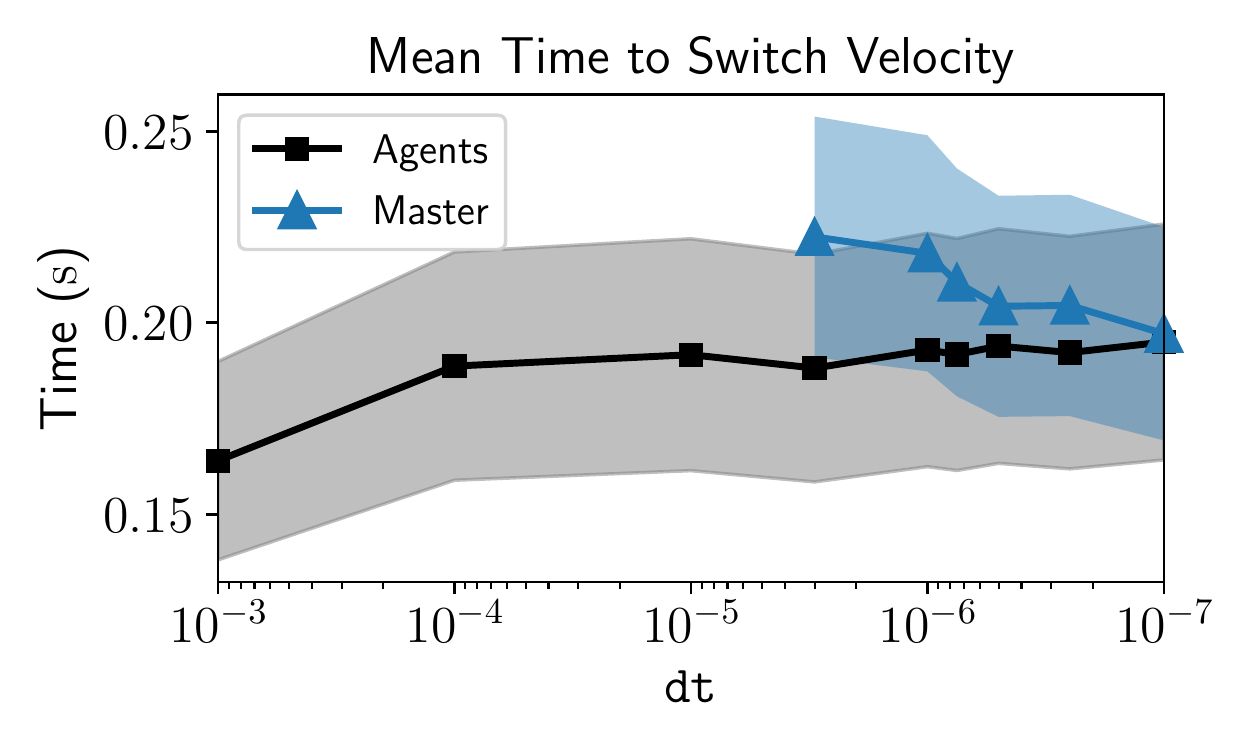}
    \caption{Mean time to switch velocity in the master equation (blue triangles) and the agent-based model (black squares). Shaded regions denote standard error of the mean ($\sigma/\sqrt{K}$, where $\sigma$ is the standard deviation, and $K$ is the number of simulations). Model parameters as in Figure \ref{fig:agents}. Agent-based model simulation parameters: \texttt{T=25} and 40 seeds for each \ttvar{dt}. Master equation simulation parameters: \texttt{T=10}, \texttt{CFL=0.3}, and 50 seeds for each \texttt{dt}. we refine the number of spatial grid points in proportion to the time step in order to maintain a constant CFL number.}\label{fig:convergence}
\end{figure}

A convergence test reveals an accurate reproduction of the mean time to switch velocity in the master equation (Figure \ref{fig:convergence}). The black line and squares demonstrate convergence in the agent-based model as a function of time step. The blue line and triangles demonstrate convergence in the master equation as a function of time step.

%We remark there generally tends to be an error between the master equation and the agent-based model on the order of \%10. 

% mfpts from fiugure 1 parameter set. see google drive file name force extension (unshared).

Note that choosing to maintain $CFL=0.3$ for all master equation simulations requires an appropriate scaling in the number of grid points (or mesh size) as a function of the time step. For example, choosing \texttt{dt=3e-6} requires \texttt{N1=N2=41} grid points to maintain $CFL=0.3$. This time step serves as an approximate upper bound, assuming we want at least 41 grid points in the mesh. We are unable to make the time step much greater because the number of grid points will decrease accordingly. This limitation comes from the choice of the extension position $B=\SI{5.05}{\nm}$, which is close to the attachment position $A=\SI{5}{\nm}$. For example, for greater extension positions such as $B=\SI{6}{\nm}$, we are allowed to choose a larger time step of size \texttt{dt=5e-5} with a corresponding number of grid points \texttt{N1=N2=50}.

\begin{figure}[ht!]
    \centering
    \includegraphics[width=1\textwidth]{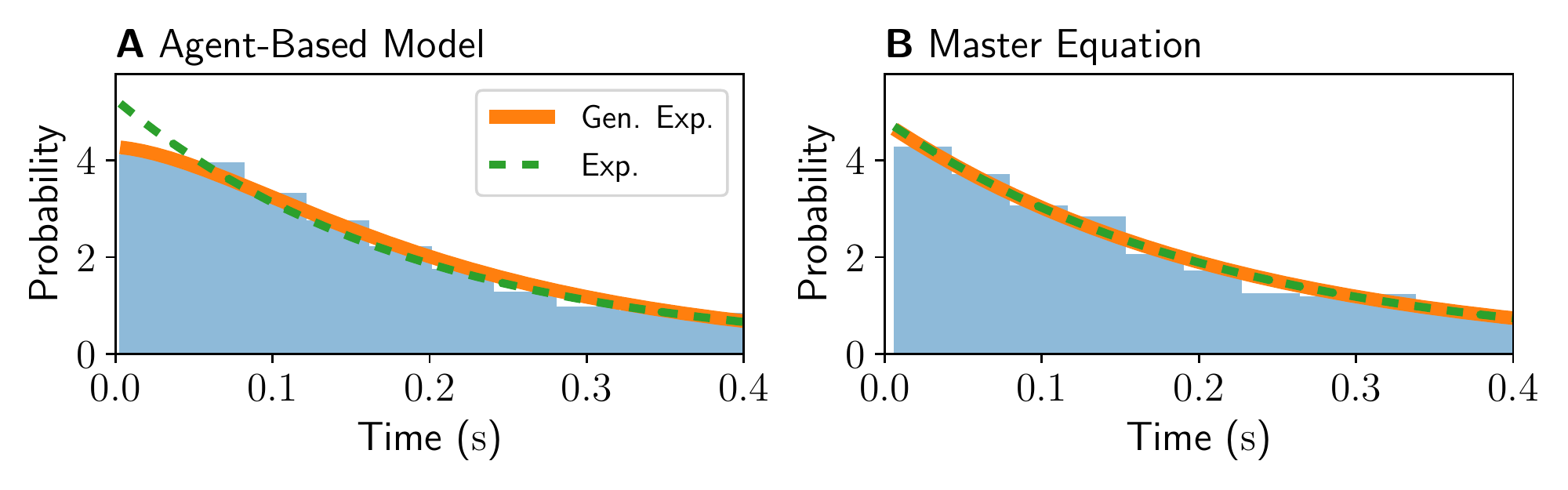}
    \caption{Waiting time distributions obtained from the agent-based model and master equation. Blue bars denote histograms of numerically computed waiting times. The orange line denotes a fit to the generalized exponential distribution, and the green dashed line denotes a fit to the exponential distribution. Agent-based model and simulation parameters are identical to Figure \ref{fig:master_v_agents}. Master equation model and simulation parameters are identical to Figure \ref{fig:master_v_agents}.}\label{fig:switch_distributions}
\end{figure}

In addition to accurate conditional mean first passage times, the distribution of waiting times is nearly identical between the agent-based model and master equation (Figure \ref{fig:switch_distributions}). The histograms satisfy either an exponential or generalized exponential distribution, defined as
\begin{equation*}
    p(x,a,b,c) = (a+b(1-\exp)) \exp\left[-ax -bx + \frac{b}{c}(1-\exp(-cx))\right],
\end{equation*}
where the three parameters $a,b,c$ are fit using a built-in maximum likelihood method. We arrive at these distributions by fitting each of the 87 distributions available in Python's \texttt{scipy} \cite{virtanen2020scipy} and quantifying the goodness-of-fit using the Kolmogorov--Smirnov (KS) test (Table \ref{tab:ks}).

Plots of the fitted exponential and generalized exponential distributions are shown in Figure \ref{fig:switch_distributions} as orange (generalized exponential) and dashed green (exponential) curves. There is no difference in fits between the generalized exponential and exponential distributions in the master equation. While the generalized exponential distribution provides the best fit to the agent-based model according to the KS test, the distribution is also  similar to an exponential distribution. Therefore, we claim the exponential distribution to be a good approximation to the master equation and move forward with this assumption. 

%https://www.tablesgenerator.com/#
\begin{table}[ht!]
    \caption{Kolmogorov--Smirnov (KS) test of proposed distributions. The exponential or generalized exponential distributions offer the best KS scores compared to all other distributions available in Python.}\label{tab:ks}
    \centering
    \begin{tabular}{l|l|l|l|l}
        & \multicolumn{2}{c|}{Agents} & \multicolumn{2}{c}{Master}  \\ \cline{2-5} 
        & Gen. Exp.      & Exp.        & Gen. Exp.     & Exp.        \\ \hline
        KS & 0.024         & 0.033  & 0.022          & 0.022           \\ 
        %$p$  & 0.6707         & 6.8e-3   & 0.760       & 2.9e-4   
    \end{tabular}
\end{table}

\section{Conditional Mean First Passage Time to Translocation}\label{sec:translocation}

In this section, we view the vesicle as a particle moving with constant velocity $\pm V^*$, where it switches to the opposite velocity after an exponentially distributed waiting time. Such a process, in which there is cross-over from ballistic motion to diffusive motion, is known as a telegraph process. To study the translocation problem, we look at this process on a finite domain $[0,L]$ with two absorbing boundaries at $z=0$ and $z=L$. Absorption at the boundary $z=L$ corresponds to successful translocation. \bluew{We remark that telegraph processes have been extensively studied in the literature, where they are referred to alternatively as persistent Brownian motion processes \cite{fan2018random,rossetto2018one}, velocity-jump processes \cite{newby2011asymptotic}, correlated random walks \cite{weiss1984first,fan2018random,tang2019first}, run-and-tumble particles (RTP) \cite{angelani2014first,angelani2015run,malakar2018steady,singh2020run}, and more generally, non-Markovian random walks \cite{hanggi1985first, masoliver1986first}. Recently, the telegraph process has seen increased interest due to its biological application to the motion of bacteria \cite{malakar2018steady,singh2020run,singh2019generalised,singh2021local, mori2020universal}}

Many existing studies on the telegraph process consider a one-dimensional domain with absorbing boundaries \cite{pinsky1991lectures,leydolt1993first}, but mean first passage time calculations often involve computing the exit time out of the entire interval \cite{weiss1984first,hanggi1985first,masoliver1986first,masoliver1992first,weiss2002some,angelani2015run,malakar2018steady,fan2018random,rossetto2018one,tang2019first,singh2020run}. In contrast, we wish to compute the mean first exit time through a particular end of the interval given an initial positive velocity and initial position $z_0\in[0,L]$. \bluew{Below, we derive the mean first exit time through the backward Fokker Planck equations developed in \cite{redner2001guide,malakar2018steady}.}

\subsection{Analytical Expressions}

To determine the conditional mean first passage time to reach $L$ given an initial condition $z_0\in[0,L]$, we recall a calculation in \cite{redner2001guide,malakar2018steady} for the probability to escape the interval through $z=L$. Let $E_\pm(z_0)$ denote the exit probability of a particle with initial position $z_0$ with positive or negative initial velocity to exit through the upper boundary at $z=L$ without touching the boundary at $z=0$. These probabilities satisfy the backward equations,
\begin{align*}
    \ypc{V^*}\pa_{z_0} E_+(z_0) - \lambda (E_+(z_0)-E_-(z_0)) &= 0,\\
    -\ypc{V^*} \pa_{z_0} E_-(z_0) + \lambda (E_+(z_0)-E_-(z_0)) &= 0,
\end{align*}
where $\lambda$ is the switching rate (see Appendix \ref{a:exit_probability} for the derivation of these equations). \ypc{To determine boundary conditions, we examine the behavior of $E_{\pm}(z_0)$ at the boundaries. With positive initial velocity at the upper boundary $z=L$, the particle escapes with unit probability. With negative initial velocity at the lower boundary $z=0$, the particle will never escape the upper boundary. The boundary conditions are therefore,
    \begin{align*}
        E_+(L) &= 1,\\
        E_{-}(0) &= 0.
\end{align*}}
\ypa{The solutions may be obtained as (e.g., using Mathematica \cite{Mathematica})}
\begin{align}
    E_+(z_0) &= \frac{V^*+\lambda  z_0}{\lambda  L+V^*},\label{eq:ep}\\
    E_-(z_0) &= \frac{\lambda  z_0}{\lambda  L+V^*}.\label{eq:em}
\end{align}
Next, let $T_\pm(z_0)$ denote the conditional mean first passage time, given an initial position $z_0$ with positive or negative initial velocity, to exit the upper boundary $z=L$ without touching the boundary at $z=0$. To compute these quantities, we define $S_\pm(z_0) \equiv E_\pm(z_0) T_\pm(z_0)$, which satisfy the equations,
\begin{align*}
    V^* \frac{d S_+(z_0)}{d z_0} - \lambda S_+(z_0) + \lambda S_{-}(z_0) &= -E_+(z_0),\\
    -V^* \frac{d S_-(z_0)}{d z_0} - \lambda S_-(z_0) + \lambda S_{+}(z_0) &= -E_-(z_0),
\end{align*}
with boundary conditions \ypc{$S_{-}(0) = E_{-}(0) T_{-}(0) = 0$ (because $E_{-}(0)=0$), and $S_+(L) = E_+(L)T_+(L) = 0$ (because $T_+(L)=0$)}. See Appendix \ref{a:exit_time} for the derivation of these equations. The solutions are,
\begin{align}
    \ypc{T_+(z_0)} &= \frac{1}{3} \left(\frac{\lambda  (L^2-z_0^2)}{\left(V^*\right)^2}+\frac{L}{\lambda  L+V^*}+\frac{2 (L-z_0)}{V^*}-\frac{z_0}{V^*+\lambda  z_0}\right),\label{eq:sp}\\
    T_-(z_0) &= \frac{1}{3} \left(\frac{\lambda  (L^2-z_0^2)}{\left(V^*\right)^2}+\frac{L}{\lambda  L+V^*}+\frac{2 L}{V^*}\right).\label{eq:sm}
\end{align}

\subsection{Numerical Translocation}

\begin{figure}[ht!]
    \centering
    \includegraphics[width=\textwidth]{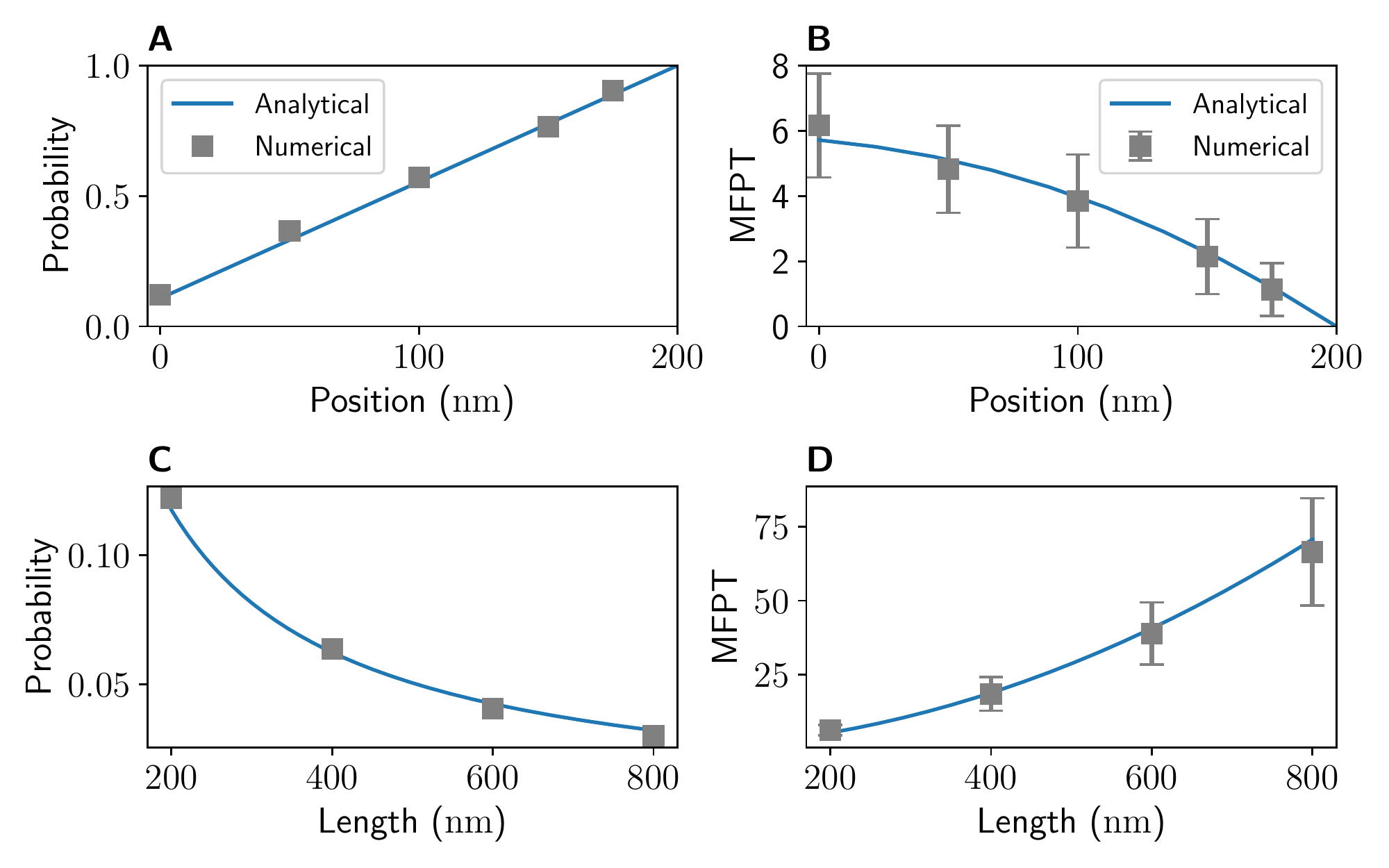}
    \caption{Probability of translocation and conditional mean first passage times with initial positive velocity. A: Probability of translocation vs. initial position $z_0$. Analytically-computed prediction (blue line) vs. the proportion of successful numerically computed translocations (gray squares). B: Conditional mean first passage time to translocation. Analytically-computed prediction (blue line) vs. numerically computed mean first passage times (gray squares). Error bars denote standard error of the mean ($\sigma/\sqrt{K}$, where $\sigma$ is the standard deviation and $K$ is the number of simulations). C: Probability of translocation as a function of spine length given initial starting position at the base of the spine ($z_0=0$). Analytically-computed prediction (blue line) vs. numerically computed mean first passage times (gray squares).  D: Mean time to translocation as a function of spine length. Analytically-computed prediction (blue line) vs. numerically computed mean first passage times (gray squares). All model parameters are identical to Figure \ref{fig:agents}. A,B: Simulation parameters for each initial position: \ypa{\texttt{CFL=0.3}, \texttt{N1=N2=41}, \texttt{dt=3e-6}, and 500 trials. C,D: Simulation parameters for each spine length: \texttt{CFL=0.3}, \texttt{N1=N2=41}, \texttt{dt=3e-6}, and 10,000 to 50,000 trials.}}\label{fig:mfpt_translocation}
\end{figure}

Spine lengths vary from \SI{0.2}{\micro\meter} to \SI{1}{\micro\meter} \cite{majewska2000regulation}. We choose a spine length of \SI{200}{\nano\meter} on the shorter end of this range for computational convenience. Shorter spine lengths mean shorter simulation times. However, we lose no generality by choosing a shorter length: the validity of approximating the translocation problem as a telegrapher's process depends on the distribution of waiting times, and the distribution is independent of spine length.

We used the master equation to run $K=500$ translocation simulations with initial positive velocity equal to the quasi-steady state velocity for this parameter set, $V^*=121$. For each representative initial position $z_0$ and for each simulation, we recorded the number of switches, time to absorption, and whether the vesicle escaped through $z=0$ or $z=L$.

To obtain the probability of escape through $z=L$, we divide the number of escapes through $z=L$ by $K$. The results are shown in Figure \ref{fig:mfpt_translocation}A, gray boxes. The numerical results are superimposed on the analytical expression $E_+(z_0)$ from \eqref{eq:ep} (blue line). To obtain the mean translocation time to escape through $z=L$, we take the average time to escape the interval $[0,L]$ conditioned on reaching $z=L$. The results are shown in Figure \ref{fig:mfpt_translocation}B, gray boxes. The numerical results are superimposed on the analytical expression $T_+(z_0)$ from \eqref{eq:sm} (blue line). Error bars denote standard error of the mean ($\sigma/\sqrt{K}$, where $\sigma$ is the standard deviation and $K$ is the number of simulations). Strong agreement between the master equation and the analytical expressions suggests that the telegraph process is a good approximation to the motor-driven vesicle dynamics.

\section{Discussion}

In summary, we use a PDE to describe the evolution of the motor position distribution to coarse-grain the microscopic analog of the mean-field vesicle trafficking equations from \cite{park2020dynamics}. While the Langevin equation appeared promising in terms of the steady-state distribution, the time to switch velocity differed greatly compared to the agent-based model. This result motivated coarse-graining using the master equation. A key component of the master equation model is the advection-reaction equation for the probability density function of motor positions. This equation allowed us to close the system, producing a coarse-grained version of the agent-based model. Finally, by noting that the waiting time to switch velocity is approximately exponential, we simplified the molecular motor dynamics as a telegraph process, which accurately predicts the conditional mean first passage time to vesicle translocation.

\ypa{The coarse-grained model performs within the same order of magnitude in wall time as the agent-based model given a relatively small number of motors. The case of small motor number is most relevant to the present study, and this is where we demonstrated the validity of the coarse-grained model in the key characteristics including bistable velocities and the mean time to switch velocities. However, we remark that the utility of the master equation is more evident in the case of greater numbers of motors. For example, with 10,000 motors, the master equation exhibits a threefold to fourfold improvement in wall time relative to the agent-based model. Such a scenario may occur when simulating motor transport of multiple cargoes in larger spaces such as axons \cite{walker2019local,kuznetsov2020modeling}.}

\ypa{Our model predictions are consistent with the experimental literature on motor-driven transport into dendritic spines in several ways. The model parameters used throughout this paper yield the quasi-stable velocity of $V^*=\SI{121}{\nm/s}$, which is close to the reported velocity range of myosin transport motors  \SIrange{200}{450}{\nm/\s} \cite{rief2000myosin}. In addition, model translocation times of \SIrange{50}{70}{\s} for spine lengths of \SIrange{500}{800}{\nm} \cite{majewska2000regulation} are consistent with the literature, where thin spines take approximately \SI{75}{\s} to translocate over \SI{1200}{\nm} and stubby spines take approximately \SI{50}{\s} to translocate over \SI{750}{\nm} \cite{da2015positioning}. 
    
Our coarse-grained model provides useful insights regarding the translocation of recycling endosomes into biological spines. In particular, it quantifies how the cell may control translocation rates by changing the shape of the spine. For example, only 2.5\% of simulated vesicles successfully translocate through an \SI{800}{\nm} spine. In general, faster switching between bistable velocities or longer spine lengths result in a decreased likelihood of translocation as shown in Equation \eqref{eq:ep} with $z_0=\SI{0}{\nm}$. This observation suggests that improvements in translocation rates may be controlled in one of two ways. First, the cell may make spines more stubby to decrease the length and reduce the confinement factor $\zeta$ (which also happens to result in less frequent switching, which in turn improves the likelihood for translocation). Second, the cell may make the spines so thin that the confinement factor dominates, resulting in a single stable positive or negative velocity depending on the dominant motor species \cite{park2020dynamics}.}

Our conditional mean first passage time calculation stands in contrast to existing studies on the telegraph process that compute the mean exit time out of an entire one-dimensional interval \cite{weiss1984first,hanggi1985first,masoliver1986first,masoliver1992first,weiss2002some,angelani2015run,malakar2018steady,fan2018random,rossetto2018one,tang2019first,singh2020run}, as opposed to computing the mean first passage time conditioned on escape through a particular endpoint. Some of these studies consider a random initial velocity with a 50/50 chance of starting with $+V^*$ or $-V^*$, whereas we condition on a positive initial velocity. The mean first passage time calculation of Bicout (1997) is closely related to our result: they calculate the mean first passage to an absorbing boundary given non-equal switching frequencies $f_+$ and $f_-$. However, their calculations include trajectories that may never return to the absorbing boundary when switching frequencies are equal. Therefore, the mean first passage time diverges in the limit $f_+=f_-$ \cite{bicout1997green}.

\ya{Note that although in the present model we assume there are multiple antagonistic motor species, in principle a tug-of-war could be possible with only a single motor species. Indeed, it is known that Myosin VI has a ``reverse gear'' and can walk bidirectionally \cite{walter2012myosin}. However, the bidirectional motion of the vesicle does not require molecular motors to be bidirectional. In fact, \cite{allard2019bidirectional} explores the existence of bidirectional motion despite using a single species of kinesin motor. In our model, informed by the biological details relevant to dendritic spines and in particular the presence of multiple motor species, we consider the up and down motors as two separate species.}

\ya{One of the assumptions of our model is that the steady-state velocities are determined by force balance between molecular motors and fluid drag. This may at first seem counterintuitive, as it is well-known that the fluid drag obtained from the Stokes drag law formula would result in a force on the order $10^{-15}$ N, i.e.~three orders of magnitude smaller than the roughly pN forces expected for molecular motors. However, carefully revisiting the assumptions of Stokes reveals that in some regimes viscous drag may become comparable to forces exerted by motors. In particular, the viscosity of cytosol can be up to 100x the viscosity of water in certain subcellular environments \cite{levitt2009membrane}. Additionally, the dendritic spine is a highly confined geometry, and estimating the effective drag in a closed constriction using lubrication theory leads to a value that is 10x higher than the free space prediction of Stokes drag law \cite{fai2017active}.}

We note that there exists a wealth of research that may be used to generalize our results to other types of molecular motors and intracellular environments. Masoliver and Weiss computed mean first passage times for a telegrapher's equation with spatially-dependent switching rates \cite{masoliver1992first}, which may allow us to incorporate biologically realistic geometries where switching rates depend on vesicle position. Additional studies examine the telegrapher's equation with asymmetric switching rates \cite{lopez2012kac}, non-equal velocities \cite{lopez2014asymmetric,rossetto2018one}, and different waiting-time distributions \cite{zacks2004generalized}, which would make it possible to incorporate kinesin and dynein motors, which are known to have distinct properties from myosin \cite{muller2008tug}. \ypa{Spatially-dependent velocities have also been considered for tau-covered microtubules \cite{newby2011asymptotic}, which may be relevant in diseased or pathologically-formed spines}. An interesting topic for future research would involve replacing the spine boundary conditions used here to consider combinations of partially reflecting, fully reflecting, and absorbing boundaries \cite{angelani2014first,angelani2015run,leydolt1993first} to simplify and incorporate questions of translocation into broader questions of spine growth and maintenance.

\bluew{Finally, we remark that, whereas for convenience our study has made use of only a few representative sets of parameters, the nonlinear model equations suggest the possibility of dramatically different behaviors in other parameter regimes. While we have performed an exhaustive search of the parameter space using the mean-field model \cite{park2020dynamics} (providing insight into the existence and stability of quasi-stable velocities as a function of constriction geometry) several quantities in the present study are unique to the stochastic model. The waiting time to switch velocity, probability to translocation, and conditional mean first passage time to translocation each depend on the model parameters and constriction geometry and can not be directly analyzed using deterministic methods. Performing a detailed exploration of the parameter space remains a promising future direction. }

\section{Acknowledgments}
We thank Kanaya Malakar and Anupam Kundu for useful discussions. The authors acknowledge support under the National Institute of Health grant T32 NS007292 (YP) and National Science Foundation grant DMS-1913093 (TGF). We acknowledge computational support from the Brandeis HPCC which is partially supported by the NSF through DMR-MRSEC 2011846 and OAC-1920147.

\appendix

\section{Pseudocode for Numerical Simulations}

\subsection{Agent-Based Model}\label{a:agents_euler}

We use an Euler scheme to simulate the agent-based model:

\begin{pseudocode}[H]
    \caption{Agent-based model. \texttt{TN} is the total number of time steps, \texttt{V} is the vesicle velocity, $f(z)$ is the force given local motor position $z$, and \texttt{$\Delta$t} is the step size.}
    \While{\texttt{i $\leq$ TN}}{
        
        \For{Each motor}{
            \eIf{Attached}{
                Increment position by $\Delta\texttt{t*V}$\;
                
                \eIf{Non-preferred direction}{
                    
                    \eIf{Position exceeds $B$}{
                        Detach\;
                    }{
                        Detach with probability \texttt{beta*dt}\;
                    }
                }{
                    Detach with probability \texttt{beta*dt}\;
                }
            }{
                Attach with probability \texttt{alpha*dt}\;
            }
            
        }
        
        \tcc{Compute attached motor forces}
        \texttt{FD}$\leftarrow$ $\sum f(\text{Down motor positions})$\;
        \texttt{FU}$\leftarrow$  $\sum f(\text{Up motor positions})$\;

        \tcc{Compute instantaneous cargo velocity}
        \texttt{V} $\leftarrow \left [ \texttt{FU} \ya{+} \texttt{FD} \right]/\zeta.$\;
        \texttt{i += 1}\;
    }
\end{pseudocode}

\subsection{Master Equation}\label{a:master_euler}

We use an Euler scheme to simulate the master equation:

\begin{pseudocode}[H]
    \caption{Master equation. \texttt{TN} is the total number of time steps, \texttt{V} is the vesicle velocity, $Q$ is the operator $Q(\phi):=-V\frac{\pa}{\pa z}\phi - \beta \phi$ and \texttt{$\Delta$t} is the step size.}
    %\KwData{this text}
    %\KwResult{how to write algorithm with \LaTeX2e }
    %initialization\;
    Initialize with D (U) down (up) motors.\;
    \While{\texttt{i $\leq$ TN}}{

        \tcc{Update the PDE of local down (up) motor positions}
        \texttt{sol[i+1]} $\leftarrow$ \texttt{sol[i] + $\Delta$t*Q(sol[i])}\;
        
        \tcc{Grow or decay according to rates in Table \ref{tab:rates}}
        D $\leftarrow$ D $\pm$ 1\;
        U $\leftarrow$ U $\pm$ 1\;
        
        Draw  D (U) positions from density \texttt{sol[i+1]}\;
        
        \tcc{Compute attached motor forces}
        \texttt{FD}$\leftarrow$ $\sum f(\text{Down motor positions})$\;
        \texttt{FU}$\leftarrow$  $\sum f(\text{Up motor positions})$\;
        
        \tcc{Compute instantaneous cargo velocity}
        \texttt{V} $\leftarrow \left [ \texttt{FU} \ya{+} \texttt{FD} \right]/\zeta.$\;
        \texttt{i += 1}\;
    }
\end{pseudocode}

\subsubsection{Upwinding Scheme for the Master Equation PDE}\label{a:upwinding}

Recall Equation \eqref{eq:advection}, the population PDE for down motors:
\begin{equation*}
	\frac{\pa}{\pa t}\phi_D\left(z,t\right) \ya{-} V\frac{\pa}{\pa z}\phi_D\left(z,t\right) = \beta \phi_D(z,t) + \alpha(1-\theta)\delta(z\ya{+}A),
\end{equation*}
We implement this equation numerically using standard upwinding/downwinding schemes depending on the sign of $V$. The upwinding scheme can be written
\begin{equation}\label{eq:upwind_step}
    \phi_i^{n+1} = \phi_i^n \ya{+} \Delta t\left[ V^+ \phi_z^- + V^- \phi_z^+ + \beta \phi_i^n\right],
\end{equation}
where $V^+=\max(V,0)$, $V^{-}=\min(V,0)$, and
\begin{equation*}
    \phi_z^- = \frac{\phi_i^n-\phi_{i-1}^n}{\Delta z_i}, \quad \phi_z^+ =\frac{\phi_{i+1}^n - \phi_i^n}{\Delta z_i},
\end{equation*}
\ypb{where \ya{$\Delta z_1 = (A_0-A)/N_1$} is the mesh size for \ya{$z> -A$} and \ya{$\Delta z_2 = (A-B)/N_2$} is the mesh size for \ya{$z\leq -A$}}. Subscripts denote position and superscripts denote time. \ypb{Let $m$ denote the index of position \ya{$z=-A$}. After computing the upwinding step \eqref{eq:upwind_step}, we integrate the source term:
    \begin{equation*}
        \phi_m^{n+1} = \phi_m^{n} + \alpha(1-\theta)/\Delta z_2.
    \end{equation*}
}

\ypa{
    \section{Derivation of Detachment Rates}\label{a:detachment_derivation}
    This section closely follows the derivation in \cite{fai2017active}. Consider down motors experiencing non-preferred velocity $V>0$. Motors attach at \ya{$z=-A$} and are stretched until they are forced to detach at \ya{$z=-B$}. Consider only those attachments with local displacement \ya{$z_0 < z < -A$}. At steady-state, it follows that
    \begin{equation*}
        \alpha(1-\theta) = \beta \ya{\int_{z_0}^{-A} \phi(z)dz} + V\phi(z_0),
    \end{equation*}
    where $\alpha$ is the attachment rate, $\theta$ is the proportion of attached down motors, $\beta$ is the detachment rate, and $\phi$ is the density of local motor positions. This equation represents a balance between the attachment of new motors at a rate $\alpha(1-\theta)$, and detachment at a rate \ya{$\beta \int_{z_0}^{-A} \phi(z)dz$} combined with flux through the position $z_0$ at a rate $V\phi(z_0)$. By taking the derivative with respect to $z_0$ and relabeling $z_0$ as $z$, we arrive at the ODE,
    \begin{equation}\label{eq:ode}
        \ya{Vd\phi/dz = \beta \phi}
    \end{equation}
    and the boundary condition $\alpha(1-\theta) = V\phi(-A)$. We can integrate this solution the set it equal to $\theta$ because the fraction of motors only occupies the interval $[A,B]$:
    \begin{equation}\label{eq:theta}
        \ya{\theta = \int_A^B \phi(z)dz = \frac{V}{\beta} \phi(-A)\{1-\exp(w)\}},
    \end{equation}
    where \ya{$w=\beta(A-B)/V$}. Combining the boundary condition along with Equations \eqref{eq:ode} and \eqref{eq:theta} yields the solution
    \begin{equation*}
       \ya{\phi(z) = \frac{\alpha \beta }{V (\alpha c + \beta)} \exp\left((z+A)\beta/V\right)},
    \end{equation*}
    where \ya{$c= 1-\exp(w)$}.
    
    We are ready to derive the expression for the detachment rate. Recall that there are two contributions to detachment: one from the basal rate $\beta$ and the other from the yield detachment rate $V\phi(B)/\theta$. We simplify this expression:
    \begin{align*}
        \ya{V\phi(-B)/\theta} &\ya{= V\phi(-B) /(V/\beta \phi(-A) (1-\exp(w)))}\\
        &\ya{= \beta (\phi(-B)/\phi(-A))/(1-\exp(w))}\\
        &\ya{= \beta \exp(w)/(1-\exp(w))}\\
        &\ya{= \beta/(\exp(-w)-1).}
    \end{align*}
    Combining the basal detachment with yield detachment results in,
    \begin{equation*}
        \ya{\beta(1+1/(\exp(-w)-1)) = \beta(\exp(-w)/(\exp(-w)-1)) = \beta/(1-\exp(w)).}
    \end{equation*}
    This is the detachment rate per motor, so that the overall detachment rate is the above expression times the number of attached motors. The derivation follows identically for up motors.}

%\section{Probability of Switching and Mean First Passage Time for Small Numbers of Switches}\label{a:small_switch}

\section{Telegraph Process}

The telegraph process is a description of the spatial probability distribution over time of a particle that moves according to two velocities $V^*$ and $-V^*$ and alternates between the velocities with a rate $\lambda$ and exponentially distributed waiting times. We consider a telegraph process on the interval $[0,L]$ with absorbing boundaries at both ends.

\subsection{Derivation of the Probability to Escape Through a Particular Interval}\label{a:exit_probability}
Let $E_\sigma(z_0,t)$ denote the exit probability of a particle with initial position $z_0$ with initial orientation $\sigma=\pm$ to exit through the upper boundary at $z=L$ at time $t$ without touching the wall at $z=0$ (to consider an initial positive velocity, take $\sigma = +$. Otherwise, $\sigma = -$). Let us consider the quantity $E_\sigma(z_0,t+\Delta t)$ \ypc{as $\Delta t \rightarrow 0$. We consider the motion in two time intervals, $[0,\Delta t]$ and $[\Delta t, t + \Delta t]$. Let us now assume that the position of the particle becomes $z_\text{new}$ after $[0,\Delta t]$. Then we have $z_\text{new} = z_0 + \sigma V^* \Delta t$. Next, we assume that the orientation of the particle becomes $\sigma_\text{new}$ after the first time interval $[0,\Delta t]$. Now $\sigma_\text{new} = -\sigma$ with probability $\lambda \Delta t$ or $\sigma_\text{new} = \sigma$ with probability $(1-\lambda \Delta t)$:
    
    \begin{equation}\label{eq:sigma_prob}
        \sigma_\text{new} = \begin{cases}
            -\sigma, & \text{with prob. } \lambda \Delta t\\
            \phantom{-}\sigma, & \text{with prob. } (1-\lambda \Delta t)
        \end{cases}.
    \end{equation}
    Now coming back to the second interval $[\Delta t, t + \Delta t]$, the particle will start with orientation $\sigma_\text{new}$ and position $z_\text{new}$ and reach the wall at $z=L$. Therefore, we have
    \begin{equation*}
        E_\sigma (z_0,t + \Delta t) = \left\langle E_{\sigma_\text{new}}(z_\text{new},t)\right\rangle_{\sigma_\text{new}}.
    \end{equation*}
    The averaging is done with respect to $\sigma_\text{new}$. Finally, using Equation \eqref{eq:sigma_prob}, we get
    \begin{align*}
        E_\sigma (z_0,t + \Delta t) &= (1-\lambda \Delta t)E_\sigma (z_\text{new}, t) + \lambda \Delta t E_{-\sigma} (z_\text{new},t)\\
        &= (1-\lambda \Dt)E_\sigma(z_0 + \sigma V^* \Dt, t) + \lambda \Dt E_{-\sigma}(z_0 - \sigma V^* \Dt, t)\\
        &= (1-\lambda \Dt)\left( E_\sigma(z_0,t) + \sigma V^* \Dt \frac{\pa E_\sigma}{\pa z_0}\right)\\
        &\quad\quad +\lambda \Dt E_{-\sigma}(z_0,t) + O(\Dt^2)\\
        &= E_\sigma(z_0,t) + \Dt\left(V^*\sigma \frac{\pa E_\sigma}{\pa z_0}- \lambda E_\sigma(z_0,t) + \lambda E_{-\sigma}(z_0,t)\right).
    \end{align*}
    Dividing both sides by $\Dt$ and taking the limit $\Dt \rightarrow 0$ yields equations for $E_\sigma$:
    \begin{equation*}
        \pa_t E_\sigma(z_0,t) = V^* \sigma \pa_{z_0} E_\sigma(z_0,t) - \lambda \left(E_\sigma(z_0,t) -  E_{-\sigma}(z_0,t)\right).
    \end{equation*}
    In the long-time limit $t\rightarrow \infty$, $E_\sigma(z_0,t) \equiv E_\sigma(z_0)$, we arrive at the desired ordinary differential equation,
    \begin{equation}\label{eq:eplus}
        V^*\sigma \frac{d }{d z_0}E_\sigma(z_0) - \lambda\left(E_\sigma(z_0) - E_{-\sigma}(z_0)\right) = 0.
    \end{equation}
    To determine boundary conditions, we examine the behavior of $E_\sigma(z_0)$ and $E_{-\sigma}(z_0)$ at the boundaries. With positive orientation at the upper boundary $z=L$, the particle escapes with unit probability. With negative orientation at the lower boundary $z=0$, the particle will never escape the upper boundary. The boundary conditions are therefore,
    \begin{align*}
        E_+(L) &= 1,\\
        E_{-}(0) &= 0.
    \end{align*}
    
}

\subsection{Derivation of the Conditional Mean First Passage Time Through a Particular Interval}\label{a:exit_time}

Let $S_\sigma(z_0,t) \equiv E_\sigma(z_0,t) T_\sigma(z_0,t)$, where $T_\sigma(z_0,t)$ denotes the conditional mean first passage time, given an initial position $z_0$ with initial orientation $\sigma$, to exit the upper boundary $z=L$ up to time $t$ without touching the boundary at $z=0$. \ypc{As above, we consider the quantity $S_\sigma(z_0,t+\Delta t)$ as $\Delta t \rightarrow 0$ in two time intervals, $[0,\Delta t]$ and $[\Delta t, t + \Delta t]$. Following the same reasoning, we arrive at
    \begin{equation*}
        S_\sigma (z_0,t + \Delta t) = \left\langle S_{\sigma_\text{new}}(z_\text{new},t)\right\rangle_{\sigma_\text{new}},
    \end{equation*}
    and using Equation \eqref{eq:sigma_prob}:
    \begin{align*}
        S_\sigma (z_0,t + \Delta t) &= (1-\lambda \Delta t)S_\sigma (z_\text{new}, t) + \lambda \Delta t S_{-\sigma} (z_\text{new},t)\\
        &=(1-\lambda \Delta t)E_\sigma(z_\text{new},t) T_\sigma(z_\text{new},t)+\lambda\Delta t E_{-\sigma}(z_\text{new},t)T_{-\sigma}(z_\text{new},t)\\
        &= (1-\lambda \Dt)E_\sigma(z_0+\sigma V^* \Delta t,t)\left\{T_\sigma(z_0 + \sigma V^* \Delta t, t) + \Delta t\right\}\\
        &\quad\quad +\lambda E_{-\sigma}(z_0-\sigma V^* \Delta t,t)\left\{T_{-\sigma}(z_0 -
        \sigma V^* \Delta t, t) + \Delta t\right\}
    \end{align*}
    Expanding in $\Delta t$, dividing by $\Delta t$, and taking the limit $\Delta t \rightarrow 0$ yields,
    \begin{equation*}
        \pa_t S_\sigma(z_0,t) = \sigma V^* \pa_{z_0} S_\sigma(z_0,t) - \lambda S_\sigma(z_0,t) + \lambda S_{-\sigma}(z_0,t) + E_\sigma(z_0,t).
    \end{equation*}
    In the long-time limit $t\rightarrow \infty$, $S_\sigma(z_0,t) \equiv S_\sigma(z_0)$, and we have the ordinary differential equation,
    \begin{equation*}
        \sigma V^* \frac{d }{d z_0}S_\sigma(z_0) - \lambda S_\sigma(z_0) + \lambda S_{-\sigma}(z_0) = -E_\sigma(z_0).
    \end{equation*}
    Combined with Equation \eqref{eq:eplus} we can solve for the conditional mean first passage time $T_\sigma(z_0)$. The boundary conditions are $S_{-}(0) = E_{-}(0) T_{-}(0) = 0$ (because $E_{-}(0)=0$), and $S_+(L) = E_+(L)T_+(L) = 0$ (because $T_+(L)=0$).}

%\bibliographystyle{plain}
%\bibliography{spines,flagella,vesicles,books,noise,fastslow,motors,code,neuro,constrictions}

\end{document}